\documentclass[twocolumn]{aastex7}
\newcommand{\kms}{\ifmmode{\,\hbox{km\,s}^{-1}}\else {\rm\,km\,s$^{-1}$}\fi}
\newcommand{\mycomment}[1]{}

\shorttitle{GD-1 and the Milky Way Starless Subhalos}
\shortauthors{Carlberg}

\begin{document}

\title{GD-1 and the Milky Way Starless Dark Matter Subhalos}

\author[0000-0002-7667-0081]{Raymond G. Carlberg}
\affiliation{Department of Astronomy \& Astrophysics,
University of Toronto,
Toronto, ON M5S 3H4, Canada} 
\email{raymond.carlberg@utoronto.ca}

\begin{abstract}
Measurements of the GD-1 star stream velocity distribution within $\pm$3 degrees of the centerline find a  total line of sight velocity spread of 5-6 \kms\ in the well measured $\phi_1=$ [-30, 0] region \citep{Valluri25}. The velocity spread is far above the $\sim$2-3 \kms\ of a dissolved globular cluster in a smooth galactic potential. Dynamical heating of the  GD-1 star stream  is simulated in an evolving model Milky Way potential which includes the subhalos extracted from cosmological CDM and WDM Milky Way-like halos.   The model bridges fully cosmological Milky Way-like halos and late time static Milky Way potentials allowing individual streams to be accurately integrated. An evolving CDM subhalo population acting for $\sim$11 Gyr heats GD-1 to $6.2\pm 1.7$ \kms. The WDM (7 keV and lighter) models develop a velocity dispersion of $3.9\pm 0.2$ \kms, only slightly greater than the 3.5 \kms\ in an evolving smooth halo without subhalos for 11 Gyr.   The dynamical age of the best model stream is close to the isochrone age of the stars in the stream.  Subhalos with masses in the decade around $10^{7.5} M_\odot$, below the mass range of dwarf galaxies, dominate the dynamical stream heating. 
\end{abstract}

\section{INTRODUCTION}

The number of globular cluster tidal star streams discovered over the last thirty-some years has increased as imaging area, sensitivity and statistical techniques have advanced \citep{Grillmair95,Odenkirchen01,g06,Malhan18,Ibata24} with now more than one hundred streams known \citep{Mateu23,BPW24}. Only recently have spectroscopic data for substantial  numbers of stream stars begun to allow metal abundance patterns and line of sight velocities to be used to identify stream stars away from the stream centerline  which is important to fully capture the stream width and velocity spread. A significant development is that the prototype thin, long stream, GD-1 \citep{GD1}, is now  recognized to have members extending  up to 3 degrees away from the centerline. The velocity distribution has a narrow $\sim $2.2-2.6 \kms\ core with velocity wings, aka cocoon, of approximately $\sim $5-8 \kms\  \citep{Ibata24,Valluri25}. These velocity widths are substantially higher than found in the velocities along the ridge line of the stream \citep{Valluri25}. A quantitative model for the development of the velocity distribution is a key challenge for stream theory.

GD-1 and the other thin tidal streams in the halo of the Milky Way have long been recognized as sensitive to perturbations from the hundreds to thousands of dark matter subhalos \citep{Ibata02,Johnston02,Carlberg09} expected to be orbiting within the Milky Way in cold and warm dark matter theories \citep{Klypin99,Moore99,Springel08}. The more massive subhalos contain visible dwarf galaxies, but below a halo mass of about $3\times 10^8 M_\odot$ \citep{Benitez-Llambay20} gas is likely not able to cool sufficiently for star formation, leaving the lower mass subhalos starless. The starless subhalos remain detectable through their gravity field, as gravitational lenses in distant galaxies \citep{Vegetti24} and dynamical effects in the Milky Way (and eventually other local group galaxies). That is, if a  $10^8 M_\odot$ subhalo having a scale radius of 1 kpc passes near or through a stream at a relative velocity of 300 \kms\  it induces a locally coherent velocity perturbation of about 1 \kms. The resulting angular momentum changes create a gap in the stream over an orbit \citep{Carlberg13,Erkal15}.  However, a stream is composed of stars on different orbits \citep{SandersBinney13} with a spread in orbital angular momentum and energy typically amounting to 2 \kms. The spread of orbital periods leads to differential motion within a stream, thereby spreading smaller gaps along the stream shortly after they form. Nevertheless, the velocity and orbital perturbations remain, gradually broadening a stream's velocity width which in turn increases the stream width.  Even though low mass halos dominate the mass spectrum of subhalos \citep{Bode01,Springel08,Lovell14},  subhalos with masses around $10^7 M_\odot$ dominate the increase of random velocities \citep{CA23,Carlberg24}.  { The increase in stream widths and random velocities originates from the same subhalo encounters that cause stream density fluctuations, so the two analyses are complementary and essentially independent measures of the integrated rate of subhalo encounters with a stream.}

It is not surprising that the globular cluster streams that develop in a cosmologically evolving  Milky Way-like halo \citep{Carlberg18,Carlberg24} have a more complex structure than those in a static late time Milky Way potential model \citep{WebbBovy19,Nibauer24}. The differences include the dynamical age, internal motions, and the  orbital spread,  all of which are larger in the cosmological simulation streams.  Overall the cosmological simulations lead to a stream population that approximates the streams in the Milky Way, but at present do not reproduce a particular stream, such as GD-1.  A more fundamental issue is that most of the subhalos that perturb a stream like GD-1 in a CDM model are starless so their cumulative effects on a stream require a statistical approach. The simulations here approximate the cosmological results to the specific orbit and progenitor star cluster of GD-1 to generate model streams that are compared to the observational data.

The goals of this paper are to further develop a dynamical understanding of stream { width and velocity} structure and to compare the results to the best available data of GD-1.   The kinematics of GD-1 member stars \citep{Valluri25} are compared to the streams formed from an n-body star particle cluster placed on the GD-1 orbit in a model Milky Way. The evolving primary potential of the model follows the mass increase of the cosmological simulation, which was designed to approximate the Milky Way.  The subhalo populations drawn from the cold or warm dark matter (CDM and WDM, respectively) cosmological simulation are integrated in the primary potential. The model streams' velocity spreads are compared to the measured GD-1 line of sight velocity dispersion to identify the model characteristics the provide the best matches. The characteristics of the observational dataset are recounted briefly in Section~\ref{sec_data}.   The straightforward standard deviation of the line of sight velocity is used to allow direct comparison with the n-body models. Section~\ref{sec_sims} describes the n-body simulations and their limitations. Section~\ref{sec_datamod} compares the observational data to the models. Section~\ref{sec_dynamics} discusses the dynamics of the models and some of their limitations. The results of the paper are discussed in Section~\ref{sec_conc}.

\section{GD-1 Data\label{sec_data}}

\begin{figure}
\includegraphics[scale=0.63,trim=10 0 0 30, clip=true]{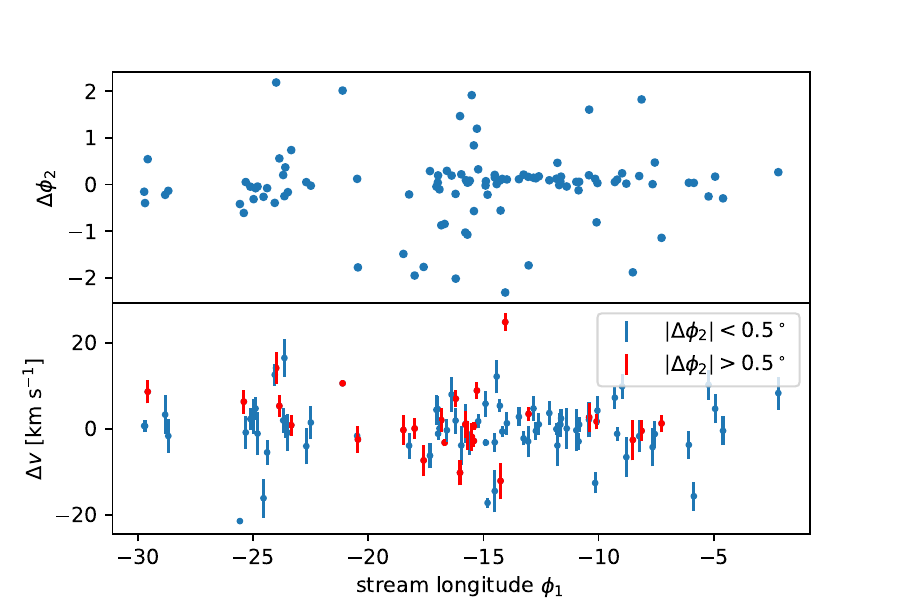}
\caption{The  $\phi_1=$ [-30, 0] segment of GD-1 on the sky, straightened with a quadratic (top panel), and the velocity offsets (bottom panel) from  \citet{Valluri25}.    }
\label{fig_dpv}
\end{figure}

The DESI-MWS survey \citep{DESI-MWS} is a component of the DESI Survey \citep{DESICollab}. The GD-1 star stream was identified from the outset as an important target that could be incorporated into the primary cosmological survey.  The DESI science verification data for GD-1 is described and analyzed in \citet{Valluri25}, hereafter V25. Figure 7 of V25 shows that the stream is best sampled in the $\phi_1=$ [-30, 0] region, which also avoids off-stream structures \citep{Bonaca19}. The data in this region was used for the V25 mixture model analysis.  Using a two component Gaussian model V25 found that the line of sight velocity distribution relative to the stream track had a core of 2.2-2.6 \kms\ and extended wings, aka cocoon, having a velocity dispersion of 5-8 \kms.   The publicly posted data of V25 has 104 stars in the [-30, 0] longitude region having formal velocity errors, $\epsilon$,  less than 5 \kms. Figure~\ref{fig_dpv} shows the data used { with quadratic fits to the mean trends removed}. The stream is rather arbitrarily divided into a core with $\left|\Delta \phi_2\right| < 0.5^\circ$ (blue) and cocoon for stars greater than $0.5^\circ$ from the centerline { to  show that there is no clear correlation between spatial and velocity offset, as shown in Figure~8 of V25}.  The root mean squared velocity error of these data, $\langle \epsilon^2\rangle^{1/2}$, is 3.05 \kms. The error corrected velocity dispersion, $\left[\langle v^2 \rangle - \langle \epsilon^2\rangle\right]^{1/2}$ is  6.0$\pm$ 0.6 \kms\ and the 3$\sigma$ clipped velocity dispersion is 5.1 $\pm$ 0.5 \kms.  Limiting the sample to  $\left|\Delta v_{los}\right| < 20$ \kms\ eliminates 2 of the 104 stars and gives the 3$\sigma$ clipped value.  One of the $>$20 \kms\ stars is a core star, the other is a cocoon star, both with relatively low velocity errors.  Allowing stars with velocity errors as large as 15 \kms\ boosts the sample size to 174, giving an error corrected velocity dispersion of 6.4 \kms. In this case the mean velocity error is 6.2 \kms, so the smaller sample with more accurate velocities having errors limited to 5 \kms\ will be used. \citet{Tavangar25} have found $\simeq5$ \kms\ velocity dispersion values in the $\phi_1=$ [-30, 0] region, with larger values elsewhere. 

The stars of the GD-1 stream of GD-1 have an [Fe/H]$\simeq$-2.5 and V25 find their color-magnitude diagram is well matched with a 12 Gyr old isochrone. Likely the progenitor cluster formed in a  pre-galactic dwarf galaxy which accreted onto the Milky Way in its early assembly, as suggested by the 14-24 kpc  inner halo orbital range of GD-1. 

\section{Simulation Setup \label{sec_sims}}

\begin{figure*}
\includegraphics[scale=0.67,trim=5 0 0 0, clip=true]{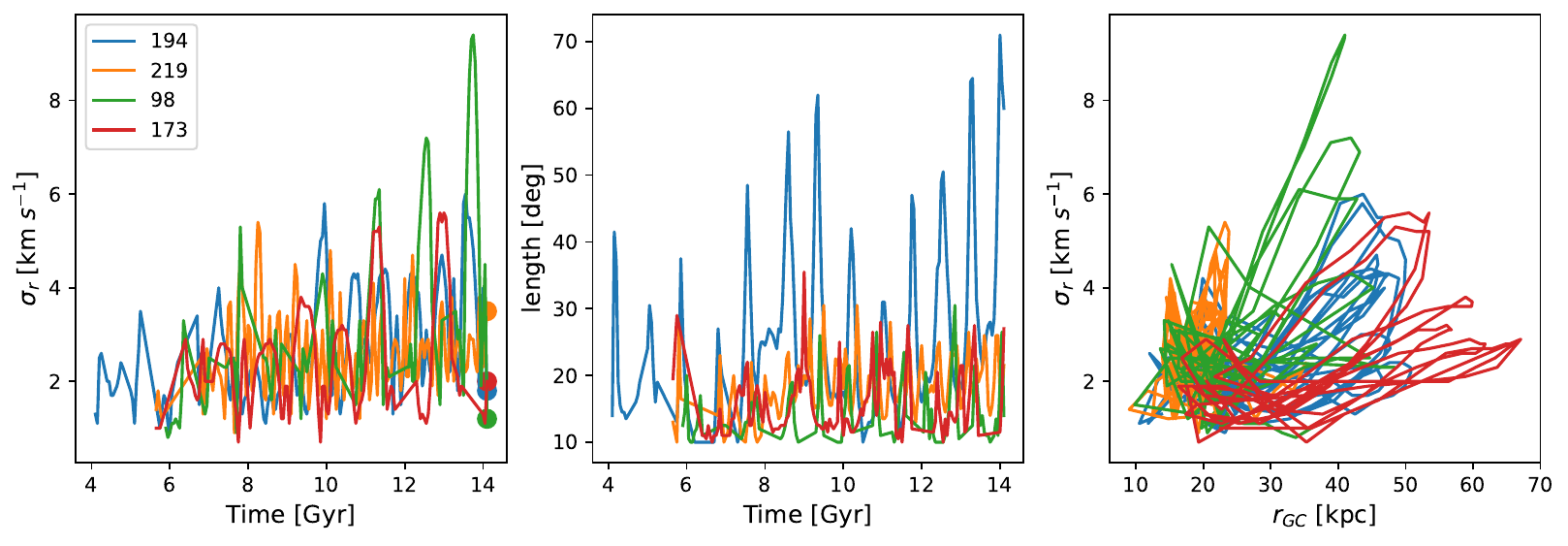}
\caption{The evolution of stream radial velocity dispersion (left panel), length (middle) and velocity dispersion with orbital position for four streams selected to have a low velocity dispersion at the end of the simulation.  }
\label{fig_siglen}
\end{figure*}

\begin{figure}
\includegraphics[scale=0.63,trim=0 50 0 70, clip=true]{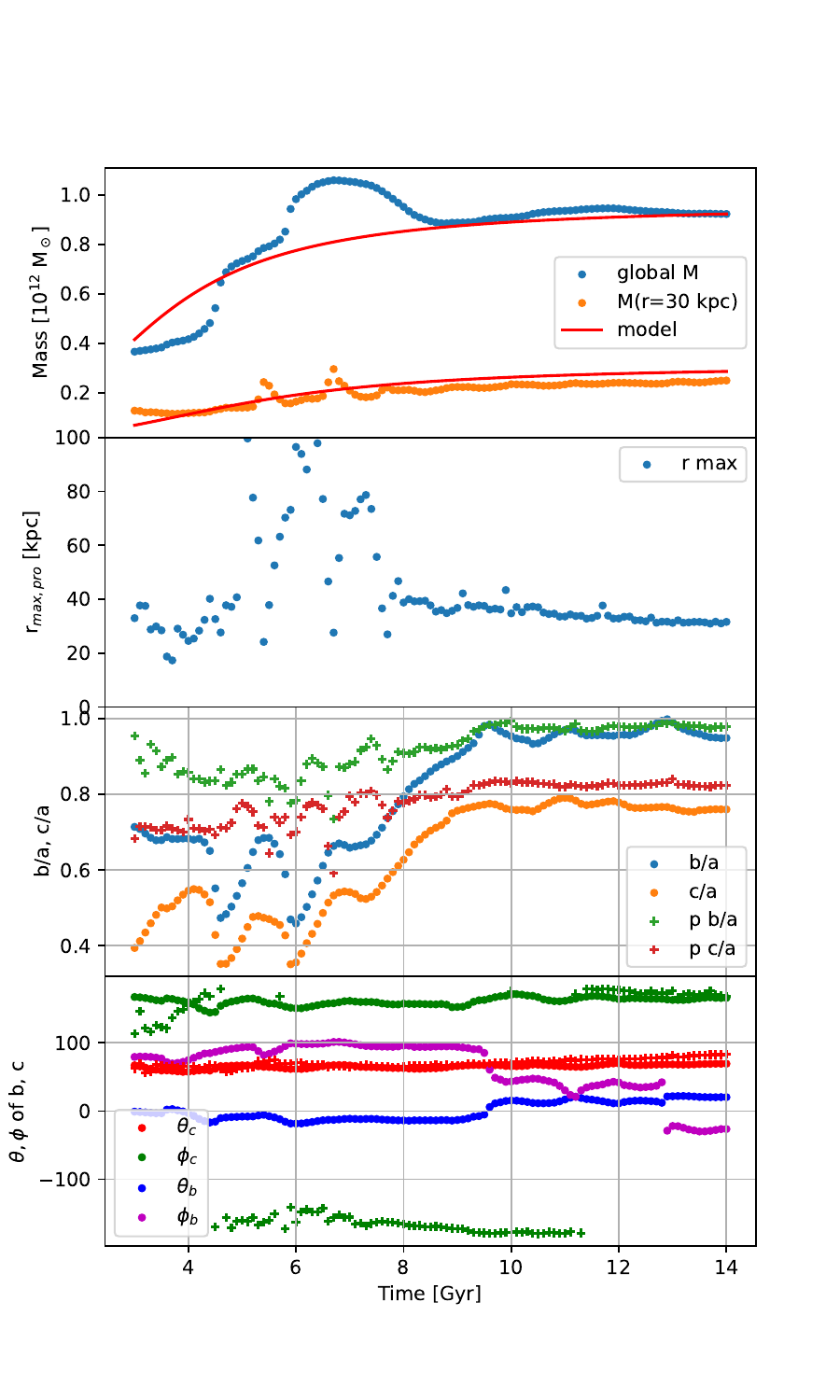}
\caption{The $M_{200}$ and mass inside a reference radius near 30 kpc (top panel), the global $r_{max}$ and the reference radius (second panel), the global triaxiality and its locally  measured at 30 kpc (third panel from top), and the orientation of the c/a and b/a axes (bottom). The simple analytic mass model with time is shown as the red line, scaled to both the $M_{200}$ mass and the mass inside 30 kpc.} 
\label{fig_Halo1}
\end{figure}

The time varying potential resuting from the hierarchical assembly of the Milky Way, $\Phi(r,t)$, is schematically the sum of an asymmetric, growing, galactic halo-bulge-disk potential, $\phi_{hbd}(r,t)$, a time-dependent set of orbiting subhalos, $\phi_{sh}(r,t)$, and occasional major mergers which distort the galactic scale potential, $\phi_{mm}(r,t)$.  The approximation is that $\Phi(r,t) \simeq  \phi_{hbd}(r,t) + \phi_{sh}(r,t)$,  where the major mergers are included in the mass increase of $\phi_{hbd}(r,t)$ and the subhalos are integrated as independent particles in the primary potential. Not included are the galactic scale asymmetries of major mergers, which cause galactic scale, $\sim$10 kpc, stream disruption, but little, small scale, $\sim$1 kpc heating. The model is then a lower limit to stream heating, expected to be accurate for the $\lesssim$10 kpc scales of interest. { Figure~\ref{fig_siglen} displays the measured velocity dispersion of the central part of four streams  from the full cosmological n-body simulation \citep{Carlberg24} selected to have a velocity dispersion of 3 \kms\ or less at the final time.    These streams, which sample a range of radii (left panel), remain coherent, although somewhat shorter (middle panel),  over the 5 and 8 Gyr time interval when the primary halo merges with two $2\times 10^{11} M_\odot$, Figure~\ref{fig_Halo1}. That is, potential fluctuations on the 10 kpc scale or less do not disrupt these streams.}

The  Milky Way-like cosmological simulations of \citet{Carlberg24} develop a primary dark halo that has a mass profile close to the Milky Way measurements \citep{Shen22} and is well fit with an NFW \citep{NFW} potential.  { Imbedded in the primary dark matter particle halo is a single particle { $5\times 10^9 M_\odot$} \citep{galpy} Plummer sphere bulge with a scale radius of 0.25 kpc. At time 5 Gyr a Miyamoto-Nagai disk \citep{MN75} centered on the bulge { in the x-y plane} begins to grow, reaching $6.8\times 10^{10} M_\odot$ \citep{galpy} at the final time. The disk and increased halo density lead to a Milky Way-like circular velocity and the density increase over a halo-only simulation increases the tidal field which erodes subhalos in the inner halo.} The AHF halo finder \citep{AHF1,AHF2} measurements of the mass within a $r_{200}$ and the mass within the region where GD-1 orbits are shown in the upper panel of  Figure~\ref{fig_Halo1}. The total mass fluctuates as mergers and accretion build up the halo, but the mass within the inner halo,  $\simeq$30 kpc, grows relatively smoothly. The second panel from the top of Figure~\ref{fig_Halo1} gives the radius of the maximum of the circular velocity and the inner halo reference radius { of 30 kpc} in which the inner halo mass of the upper panel is measured.  The variation of the primary halo mass with time is approximated with the function $\arctan{[(0.3 t/{\rm Gyr})^2]}$ normalized to produce the final halo mass of $9.225\times 10^{11} M_\odot$ which is a good approximation to the halo mass inside 30 kpc. The halo mass profile is fit with an NFW \citep{NFW} function with $c=13.8$  and $r_s = 20$ kpc, with $r_s$ kept constant as Figure~\ref{fig_Halo1} indicates. The final time potential shape parameters are estimated to be $q_b=0.97$ and $q_c=0.83$ which are aligned with the y and z axes, respectively, as an approximation to the relatively slowly varying halo orientation angle measurements of Figure~\ref{fig_Halo1}.  { AHF calculates shape and orientation from the moment of inertia of particles within the specified radius.}

The subhalos in the cosmological simulations \citep{Carlberg24} are identified with the AHF halo finder in redshift zero physical coordinates and densities. Halos with masses above $1\times 10^6 M_\odot$ and below $3\times 10^8 M_\odot$ are { approximated as Hernquist spheres, using their measured $r_{max}=2a$ to calculate the scale radius, $a$}, placed into the evolving model potential at their positions and velocities and integrated forward. To allow for the reduction in subhalo numbers with time in the virialized halo \citep{Carlberg24} the masses of {  all the subhalos are uniformly} reduced with an exponential decline beginning at 5 Gyr with a time constant of 5.6 Gyr,  the time decay measured in the cosmological simulations. { The scale radii are reduced in proportion to the subhalo mass to the 1/3 power to keep the subhalo density constant.}  The mass decline with the { cumulative mass-number relation $N(>M) \propto M^{-1}$ for CDM and the massive subhalos of WDM}  ensures that that the numbers at a given mass decline as required. { The model predicts 116 subhalos above $10^7 M_\odot$ within 100 kpc at the final time whereas the cosmological n-body simulation finds 105.} The known dwarf galaxies \citep{McConnachie12} are added with halo maximal halo masses, $\log_{10}{(M_h)} = \log_{10}{(5 L_V)}+0.65[8-\log_{10}{(L_V)}]$ which approximately abundance matches them to the subhalos above $3\times 10^8 M_\odot$ of the simulation. The 5.5 keV WDM subhalos { below a density of $3.6 \times 10^6 M_\odot {\rm kpc}^{-1}$  are removed because the low density early time WDM subhalos \citep{Lovell14}  do not survive in the high density inner region that develops in the primary halo beyond 7 Gyr age}. No density cut is required for the 7 keV WDM subhalos. 

The star clusters are set up with a King model, use $1 M_\odot$ star particles with a 1 pc softening. The clusters have initial half mass radii of 3-5 pc which sets the half mass radius density, which determines the relaxation time, which controls the rate at which the cluster loses stars into the tidal region where they are heated and swept away.  { The two-body effects are added with a Monte Carlo heating of stars inside the half mass radius, calibrated to an N-body6 cluster \citep{Carlberg18}. The tidal radius of the clusters vary between 50 and 100 pc so the orbits beyond the cluster half mass radius of 5pc are insensitive to the softening and Monte Carlo heating.}The nominal final time location of the center of the dissolved progenitor star cluster uses the \citet{WebbBovy19} values in the \citet{Koposov10,PW17} frame which puts the progenitor at a stream longitude of $\phi_{1} =-40$. These orbital values generate a late time orbit that is a good but not unique match to the data.  The initial conditions are backward integrated in the evolving main potential along with the evolving and orbiting subhalos to find the star cluster starting time coordinates. The n-body model for the star cluster uses the subhalos and time varying main halo potential as pre-computed external forces. The small differences in orbital time steps between runs in the complex potential causes the progenitor cluster to come back to slightly different final time positions. The  spread of final positions increases as the  subhalo numbers rise and with the time length of the integration. The times for simulation are kpc/\kms, referred to as simulation Gyr, and are 1.022 true Gyr. Movies showing the time evolution of the streams, the subhalos and subhalos encounters with stream particles are at \href{https://www.astro.utoronto.ca/~carlberg/streams/gd1}{GD-1 Stream Simulations}.

\section{Simulated Streams in an Evolving Potential\label{sec_datamod}}

\begin{figure}
\includegraphics[scale=0.23,trim=0 0 10 0, clip=true]{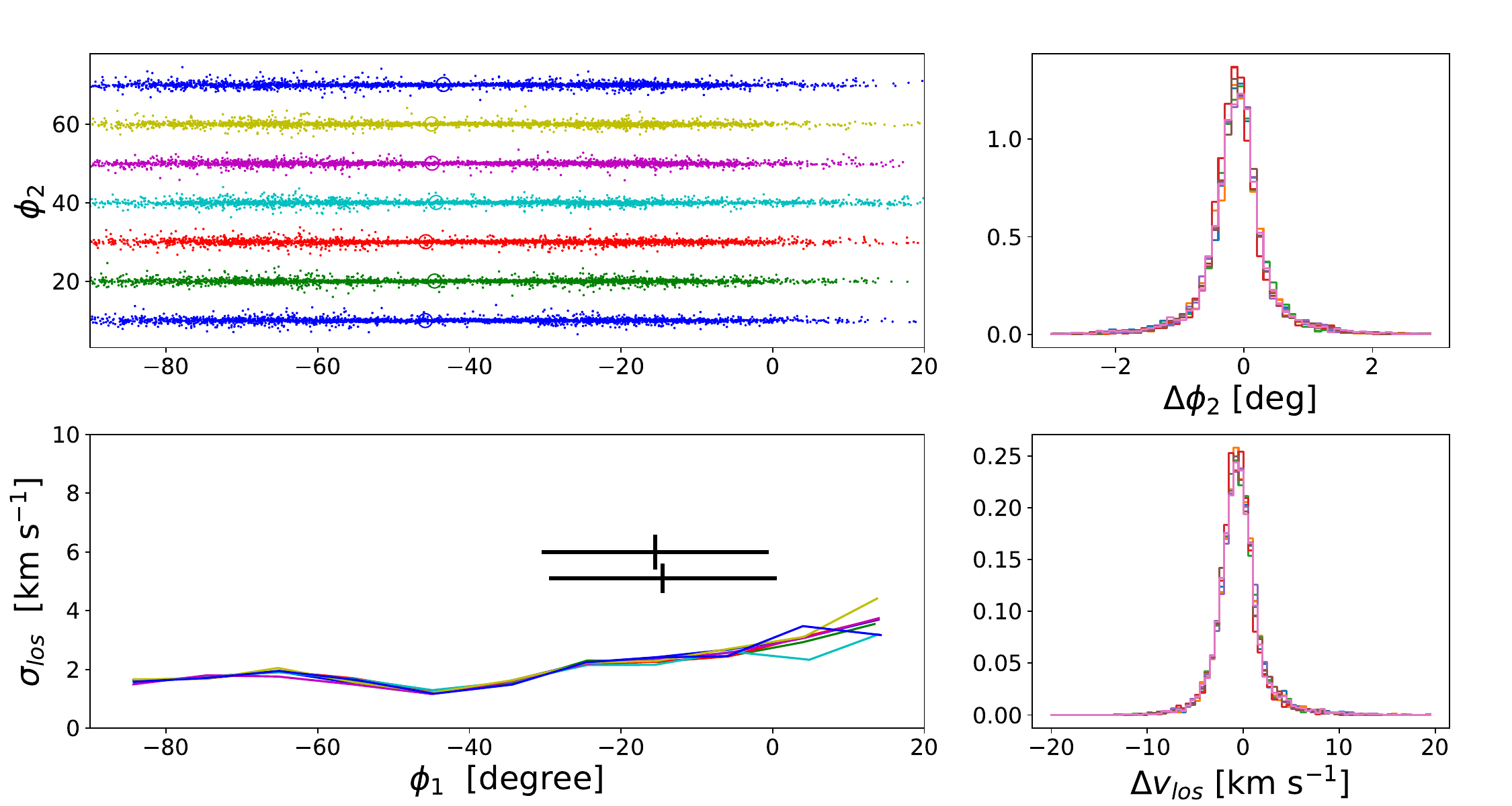}
\includegraphics[scale=0.23,trim=0 0 10 0, clip=true]{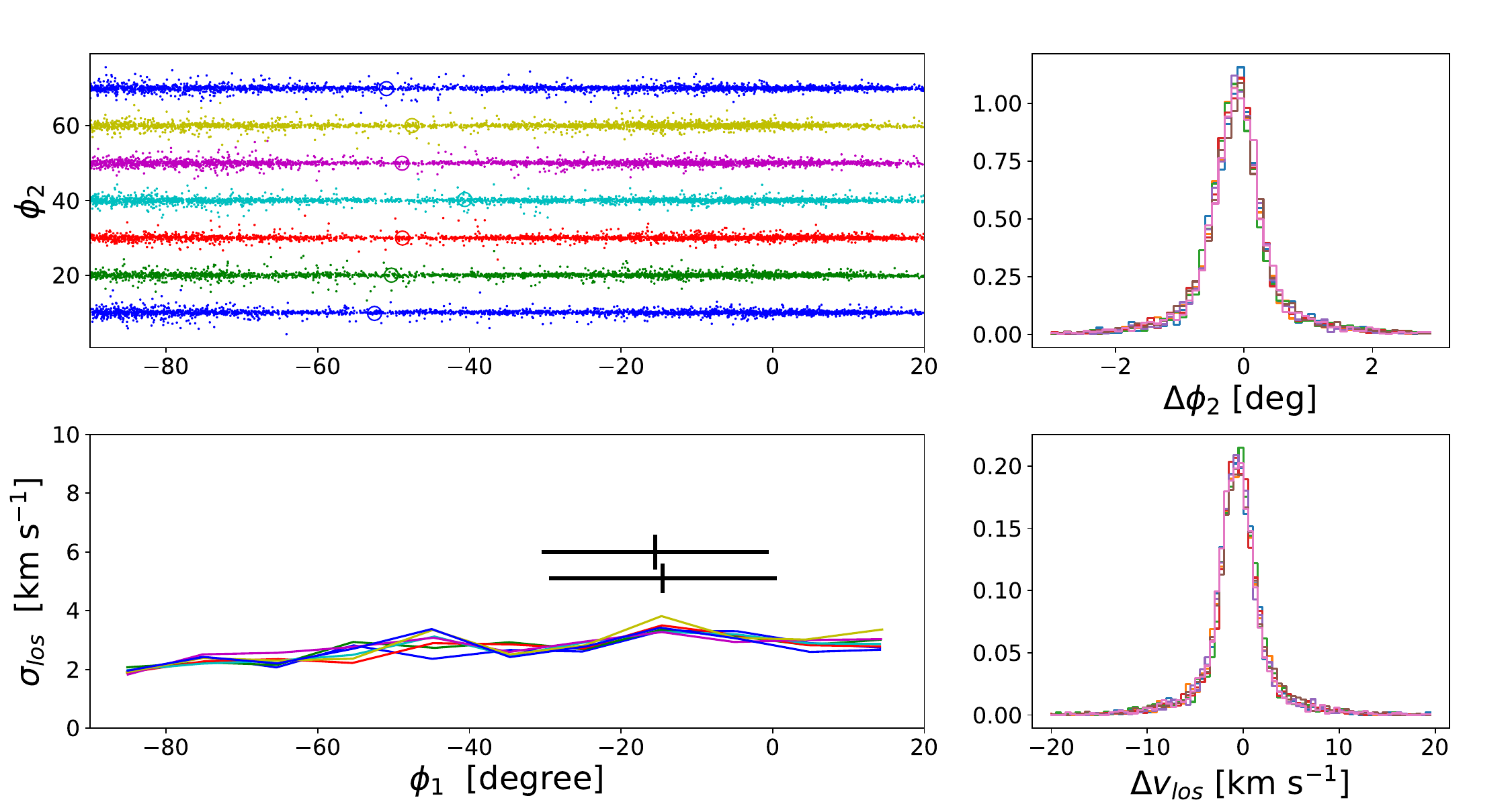}
\includegraphics[scale=0.23,trim=0 0 10 0, clip=true]{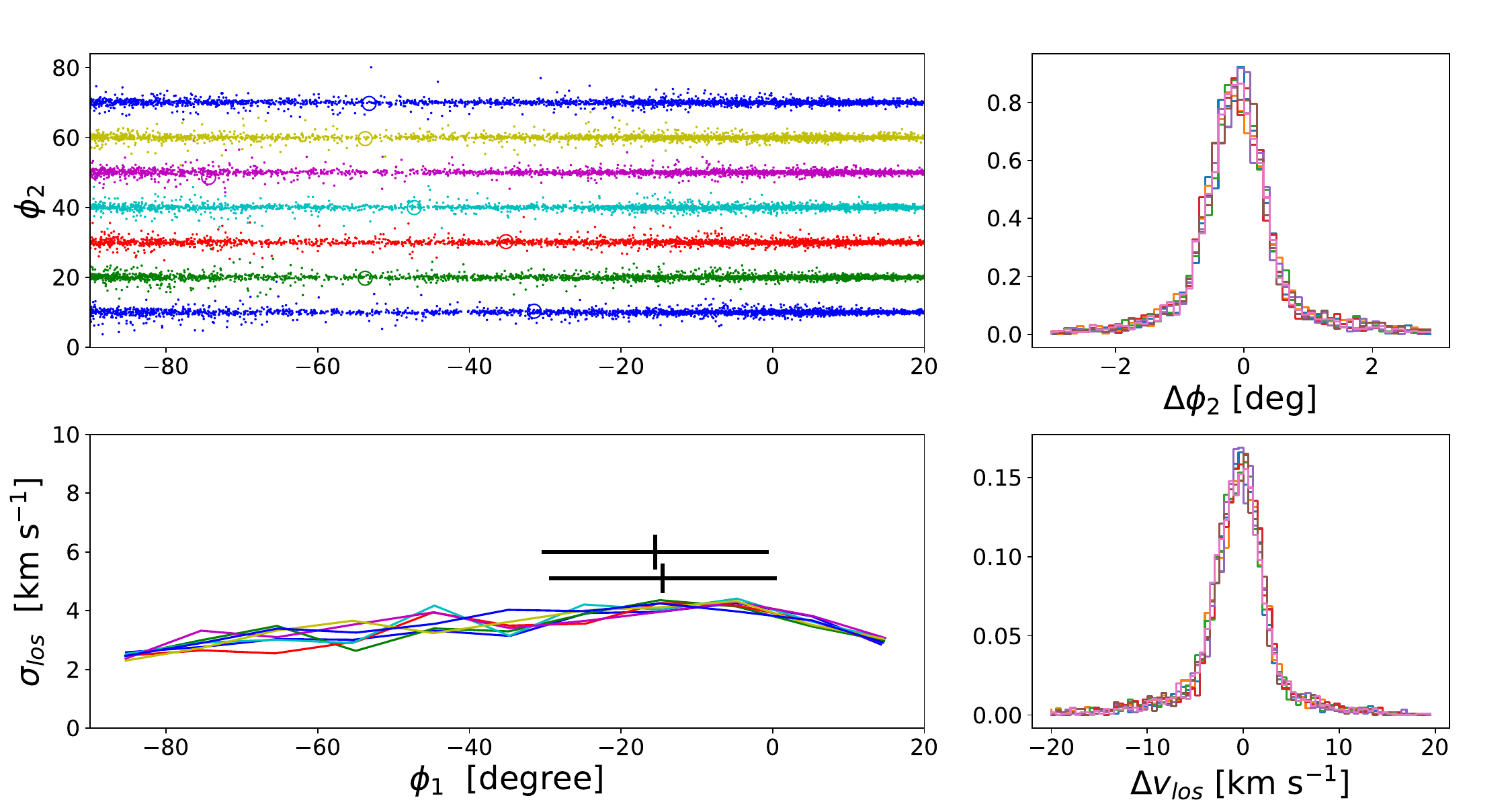}
\includegraphics[scale=0.23,trim=0 0 10 0, clip=true]{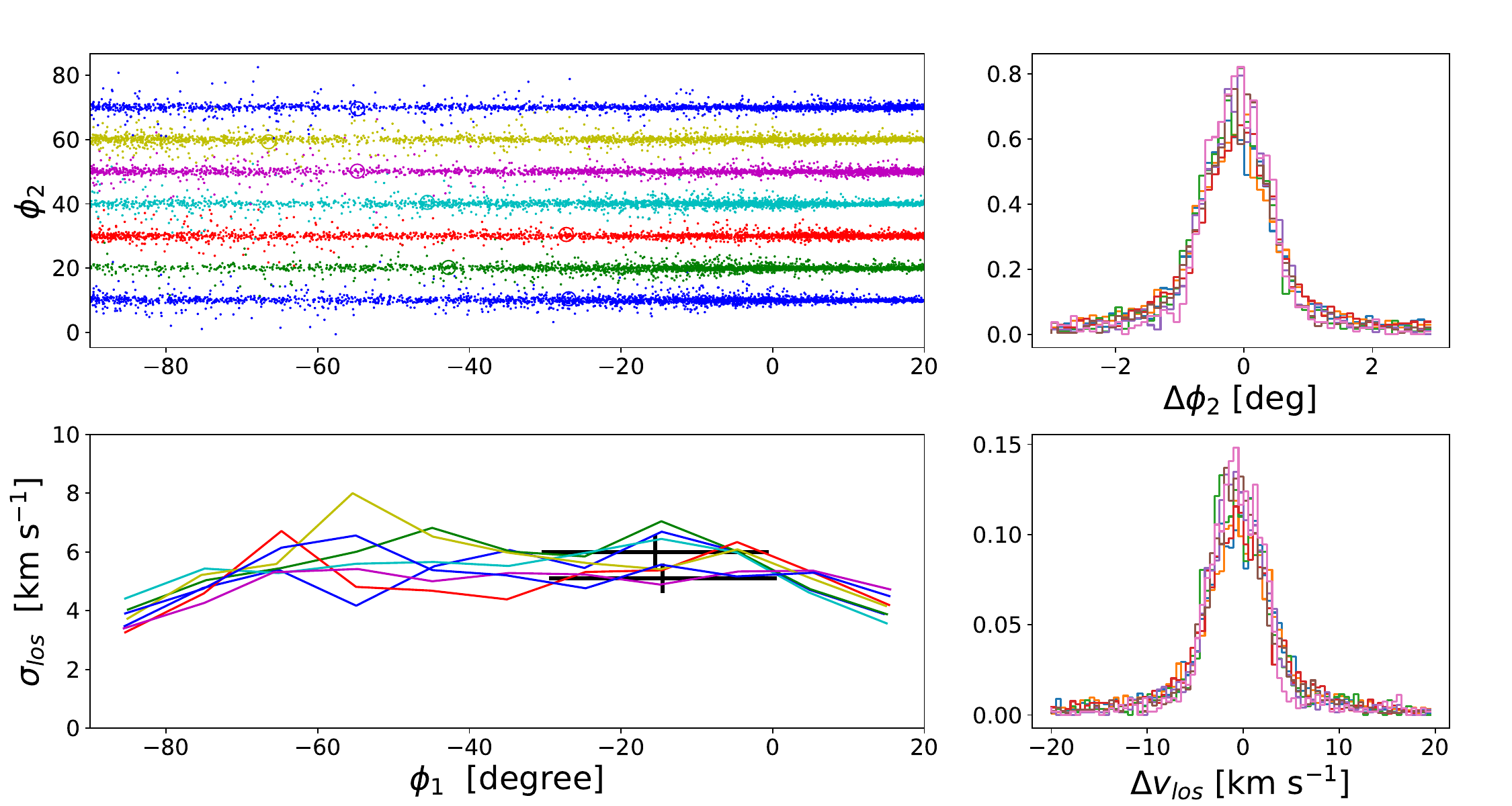}
\caption{Stream models at 3 (top), 5, 7 and 10 (bottom) Gyr with a $2\times 10^4 M_\odot$ progenitor. The black error bars are the measured velocity dispersions over the $\phi_1=$ [-22,-4] range of the error bar. The colored lines are for the 6 rotations of 30 degrees of the subhalos with $2\sigma$ (dashed) and all velocities (solid).  The histograms on the right sum the data along the streams and summarize the model velocities which are presented in Table~\ref{tab_sigt}.
} 
\label{fig_cdm_age}
\end{figure}

Figure~\ref{fig_cdm_age} shows the streams resulting from a $2\times 10^4 M_\odot$ progenitor \citep{Koposov10} (colored dots and lines) with $\Delta \phi_2$ (upper panel pair) and $\Delta v_{los}$ (lower panel pair) for a sequence of stream ages. Each stream age is simulated six times with the subhalo distribution rotated 30 degrees between each run. The mean $\phi_2$ and $v_{los}$ dependence with $\phi_1$ is fitted with a fourth order polynomial and subtracted to give the $\Delta$ differences from the mean trend. { The results are insensitive to any fit between a quadratic and a sixth order fit.} The black error bar is the velocity dispersion derived from the V25 data.  The 20,000 star particles are randomly sampled down to 1000 particles to create plots with particle numbers comparable to the observable number of stars \citep{Tavangar25}.  The colored circle shows the final time position of the tracer particle at the center of the now dissolved progenitor cluster. The upper right panel shows the  width distribution around the mean. The standard deviation widths are given within the panel. The lower left panel shows the standard deviation of the velocity measurements of the particles (colored lines).  The lower right panel shows the velocity distribution. 

\begin{table} 
\caption{$\sigma_{los}$, $2\times 10^4 M_\odot$ }
\label{tab_sigt}
 \begin{center}
 \hskip-50pt \begin{tabular}{|r|r|r|r|}
 \hline
 age & \multicolumn{2}{c|}{$\sigma_{los}$, $\phi_1=[-30,0]$} &  M [-90, 20]  \\
 \hline
 ~& all &  $\left| \Delta v \right| <3\sigma$  &  ~~\\
\hline
  Gyr   & \multicolumn{2}{c|}{\kms} & $M_\odot$ \\
 \hline
3 & 2.33 & 1.76 & 19472\\
5 & 3.33 & 2.14 & 15851\\
7 & 4.25 & 2.64 & 12272\\
9 & 4.97 & 2.90 & 11988\\
10 &6.21 & 3.59 &9581\\
\hline
\end{tabular}
\end{center} 
\end{table}

Table~\ref{tab_sigt} shows that model streams 9 Gyr and younger develop a smaller velocity width (1.6$\sigma$ confidence) than the $6\pm0.6$ \kms\ of GD-1.  A 10 Gyr age stream is in approximate agreement with the data (lowest panels pairs), however the mass within a generously defined visible portion of the stream, $\phi_1=$[-90, 20] is 9581 $M_\odot$, below even the lower end mass estimates \citep{Tavangar25}.  The heating of the stream rises rapidly beyond 7 Gyr stream age (simulation times 7 Gyr and less) because there are more subhalos at early times. 

\begin{figure}
\includegraphics[scale=0.23,trim=0 0 10 0, clip=true]{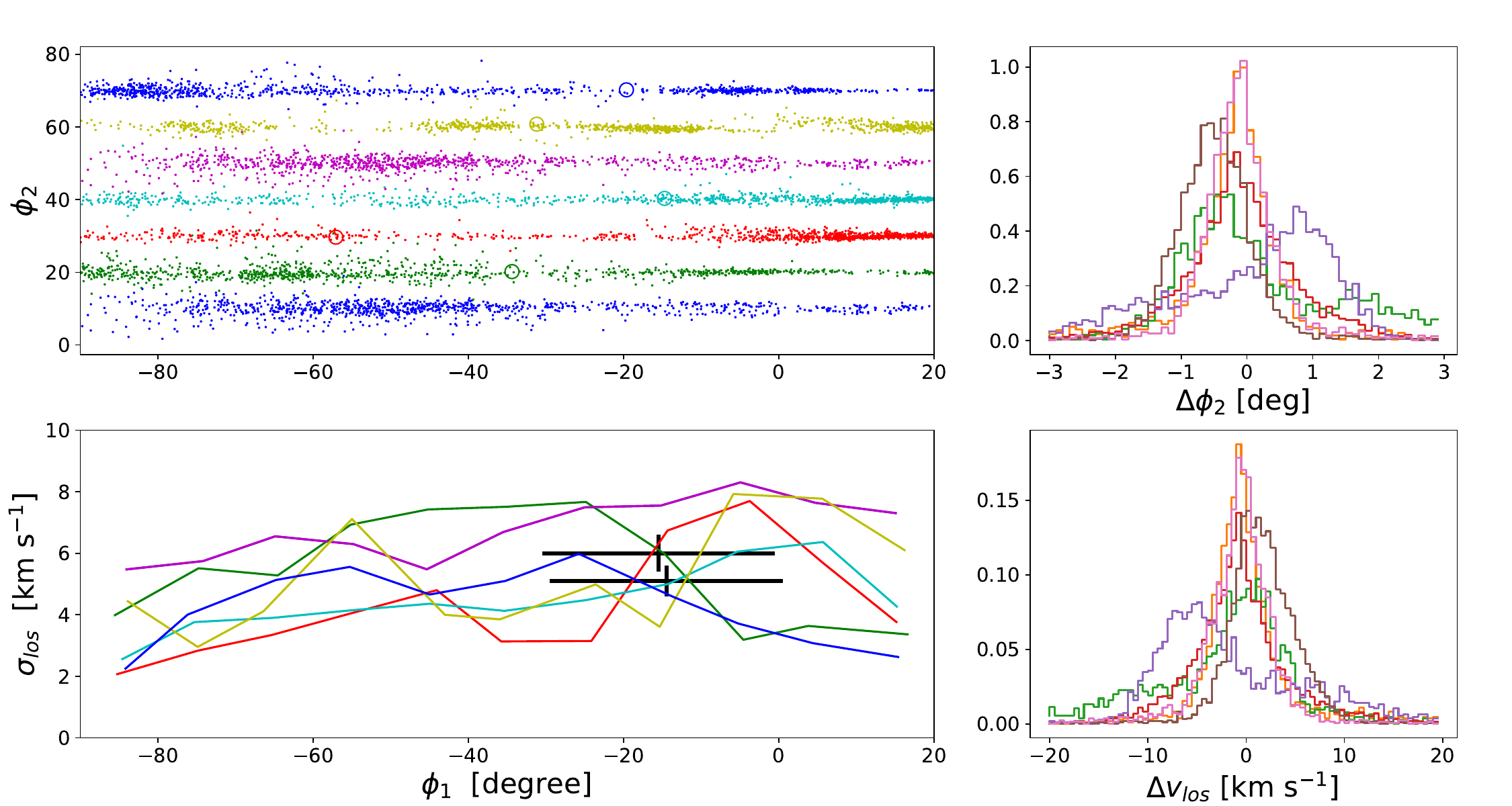}
\includegraphics[scale=0.23,trim=0 0 10 0, clip=true]{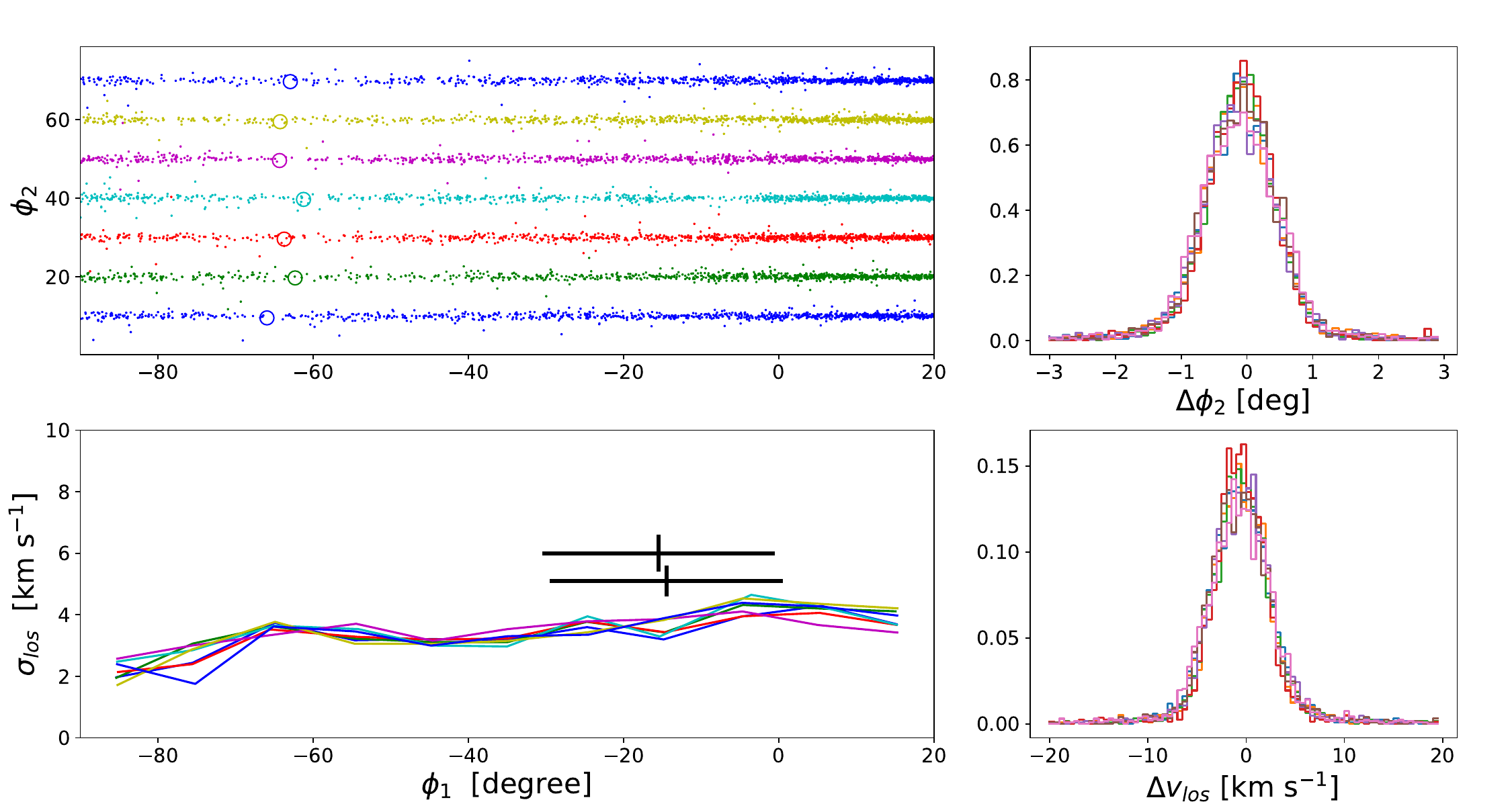}
\caption{Stream models for a  $5\times 10^4 M_\odot$ progenitor in CDM (top) and WDM(7 keV) for 11 Gyr. Velocity spreads are in Table~\ref{tab_siglos}.}
\label{fig_cdm11}
\end{figure}

A more massive progenitor of $5\times 10^4 M_\odot$ \citep{Ibata24} is used for Figure~\ref{fig_cdm11} to boost the visible stream mass. This progenitor leaves 20419 $M_\odot$ in the visible section of GD-1. The progenitor star particle cluster  is started with an initial half mass  radius of 5.5 pc, rather than 3 pc for the $2\times 10^4 M_\odot$, making them less dense which leads mass halving time of $\sim$1.5 Gyr, rather than $\sim$0.5 Gyr for the $2\times 10^4 M_\odot$ mass progenitors. A longer dissolution time means that new stream members are added later in an environment with fewer subhalos, sharpening the core-wing structure of the stream.

Models of GD-1 have generally used stream ages in the 3-5 Gyr range on the basis that older streams become longer than the apparent length of GD-1 \citep{WebbBovy19,Nibauer24}. As discussed briefly in \citet{Carlberg24} and readily visible in the stream movies  \href{https://www.astro.utoronto.ca/~carlberg/streams}{Streams Data} streams within 30 kpc of the galactic center are usually completely wrapped around the galaxy, with a central region at high surface density on the sky but with the more distant stars at such low sky density and large velocity spread they are much smaller fraction of the field population \citep{}.  In any range of distance and proper motion they are hard to find with current photo-kinematic stream finders. The movies show that the central region of a stream stays coherent at high sky density but more distant particles are taken on  divergent orbits as larger scale potential fluctuations occur from time to time.

\begin{figure}
\includegraphics[scale=0.48,trim=20 0 20 20, clip=true]{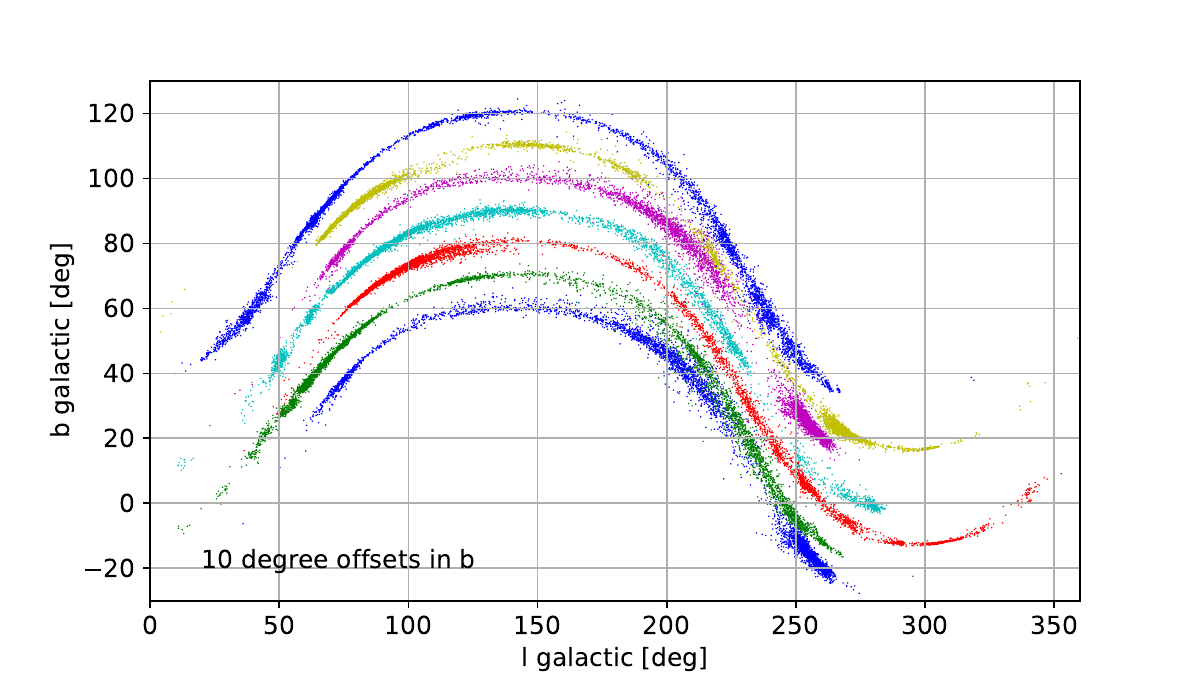}
\caption{The 11 Gyr CDM models of Figure~\ref{fig_cdm11} plotted in galactic coordinates. Each successive stream from the bottom has a 10 degree latitude offset added. A random selection of 2000 star particles of the 50000 simulated are plotted.}
\label{fig_galactic}
\end{figure}

The models with a $5\times 10^4 M_\odot$ progenitor run for 11 Gyr are shown in galactic coordinates in Figure~\ref{fig_galactic}. The currently known stream stretches from approximately galactic longitude 90 degrees to 240 degrees \citep{Ibata24}. The southern hemisphere stream track goes as far south as -35 galactic latitude, but not all stream models populate the track that far south. Both ends of the northern track GD-1 are somewhat poorly defined as they enter regions of high extinction below galactic latitude 30 degrees. About half of the mass of the models is within 20 degrees of the plane where it is obscured. The stream ends in the CDM models are highly variable. Furthermore  alternate progenitor locations along its orbit and a wider range of progenitor masses and radii are not explored. The more important consideration for the large scale behavior of the stream is the evolving potential itself. The model here has the primary halo and the subhalos but does not include the merging halos on intermediate mass scales which causes the older parts of the stream to follow a diverging orbit as is readily apparent in the movies at  \href{https://www.astro.utoronto.ca/~carlberg/streams/417}{CDM Low Mass Cluster Streams}. 

\begin{table} 
\caption{$\sigma_{los}$,  $5\times 10^4 M_\odot$ }
\label{tab_siglos}
 \begin{center}
  \begin{tabular}{|r|r|r|r|} 
 \hline
stream & \multicolumn{2}{c|}{$\sigma_{los}$, $\phi_1=[-30,0]$}  &  M [-90, 20]  \\
 \hline
 ~& all & $\left| \Delta v \right| <3\sigma$ & $M_\odot$ \\
\hline
11 Gyr   & \multicolumn{2}{c|}{\kms} & \\
 \hline
CDM & 6.22 & 4.80 & 20419\\
WDM 7 keV & 3.93 & 2.76 & 13580\\
no subhalos & 3.49 & 2.62 & 11413\\
\hline
10 Gyr   & \multicolumn{2}{c|}{\kms} & \\
\hline
CDM & 4.00 & 2.86 & 16870 \\
WDM 7 keV &3.62 & 2.69 & 15597 \\
\hline
GD-1 ($\Delta v < 30$) & 6.0 $\pm$ 0.6& 5.1 $\pm$ 0.5& $\sim$20000 \\
GD-1 ($\Delta v < 20$) & 5.1 $\pm$ 0.5& 5.1 $\pm$ 0.5& ... \\
\hline
\end{tabular}
\end{center}
\end{table}

 \section{Dynamical Interpretation\label{sec_dynamics}}
 
{ The velocity distribution of the member stars of the  GD-1 stream has wings wider than a Gaussian \citep{Malhan19,Ibata24,Valluri25}, although the result depends on accurately identifying stream members}.  There are several ideas for the origin of the wings, aka cocoon, as considered in V25. This paper shows that stars can be perturbed into the wings through dynamical interactions with the starless subhalos present in the galactic halo. One consequence is that the overall velocity dispersion of the stream is then the total velocity spread, not the core velocity dispersion, which means that the blurring of the perturbations from individual sub-halos, that is gaps, is proportionally larger. The data are consistent with the larger velocity dispersion, dynamical heating view, but more data on more streams are important to bolster this view.
 
\begin{figure}
\includegraphics[scale=0.75,trim=0 0 0 20, clip=true]{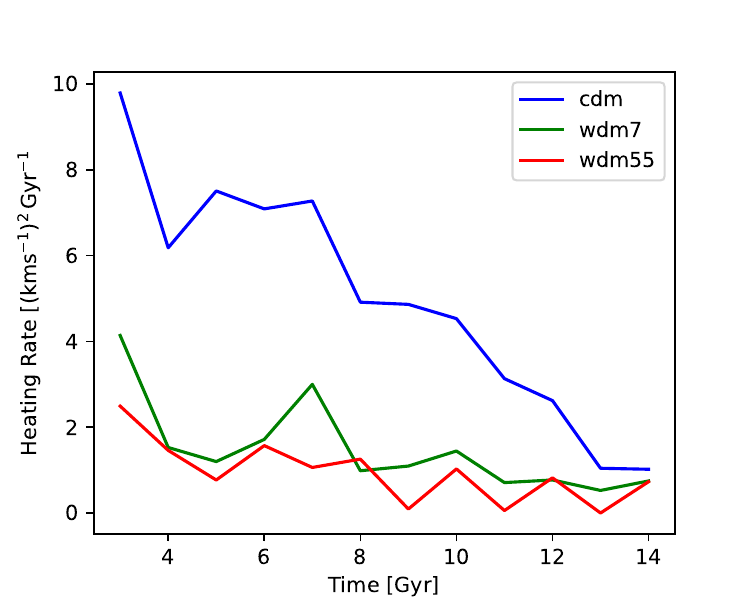}
\caption{The time dependence of the random walk heating rate,  $d \langle \delta v^2 \rangle /dt$, of a stream  from the subhalos found with time in the 10-30 kpc region of three cosmological n-body simulations.}
\label{fig_heating}
\end{figure}

A model of stream heating via subhalo stream crossings is developed in \citet{CA23,Carlberg24}. The impact approximation gives the velocity perturbations $\delta v(M)$ that a subhalo of mass $M$ causes in the stream. The velocity perturbations are treated as a random walk and summed over the mass spectrum of the subhalos to calculate a heating rate, $d \langle [\delta v(t)]^2 \rangle/dt$. The rate of encounters is steeply falling function of subhalo mass, $dN/dM \sim M^{-1.9}$ in CDM, with the encounters becoming less than one per 4~Gyr at a mass of $\sim 10^{7.5} M_\odot$, where the heating summed over masses is truncated. The heating rate calculation uses the impact approximation to calculate the velocity perturbations for  the subhalos found in the cosmological n-body simulations between 10 and 30 kpc measured once per Gyr, for the $\simeq$300 \kms\ average encounter speed on a 10 kpc stream, as in \citet{Carlberg24}.  The alternate approach of summing the $\delta v$ of individual encounters in the evolving potential model for the GD-1 stream gives a similar summed velocity dispersion of about 9 \kms. The  CDM heating rate of  Figure~\ref{fig_heating} is much higher at early times because there are far more subhalos then, a consequence of the Milky Way-like system having a declining accretion rate and no major mergers after 7 Gyr, and the development of a massive disk that causes  approximately half of the subhalos in the inner region of the galaxy to dissolve relative to a no-disk halo \citep{Donghia10,GKBullock17,CA23}.

\begin{figure}
\includegraphics[scale=0.75,trim=0 0 0 20, clip=true]{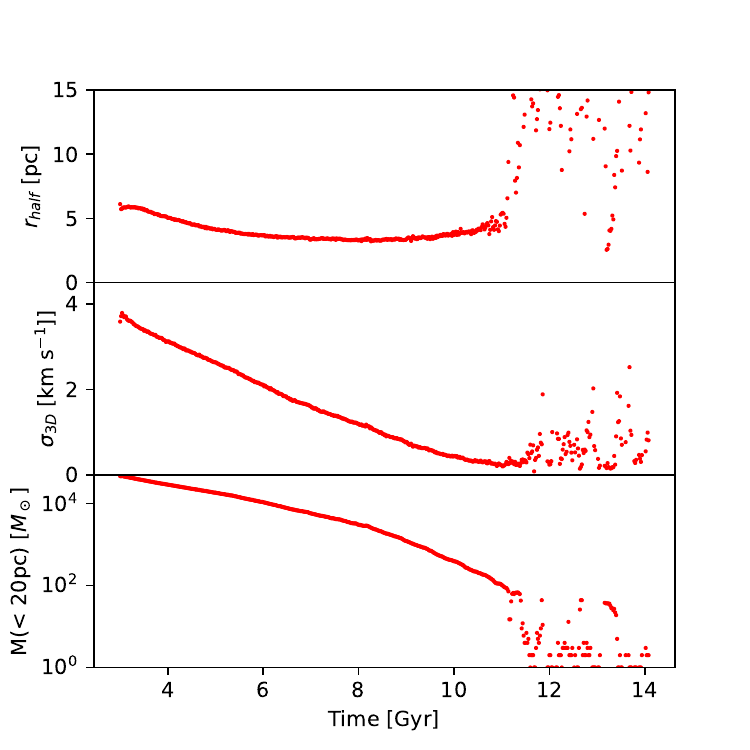}
\caption{The half-mass radius, 3D velocity dispersion and mass with time of a $5\times 10^4 M_\odot$ progenitor cluster in a CDM subhalo background. The cluster is effectively unbound at time 11.5 Gyr but a few particles continue to orbit together.}
\label{fig_cluster}
\end{figure}

\begin{figure*}
\includegraphics[scale=0.6,trim=60 20 50 20, clip=true]{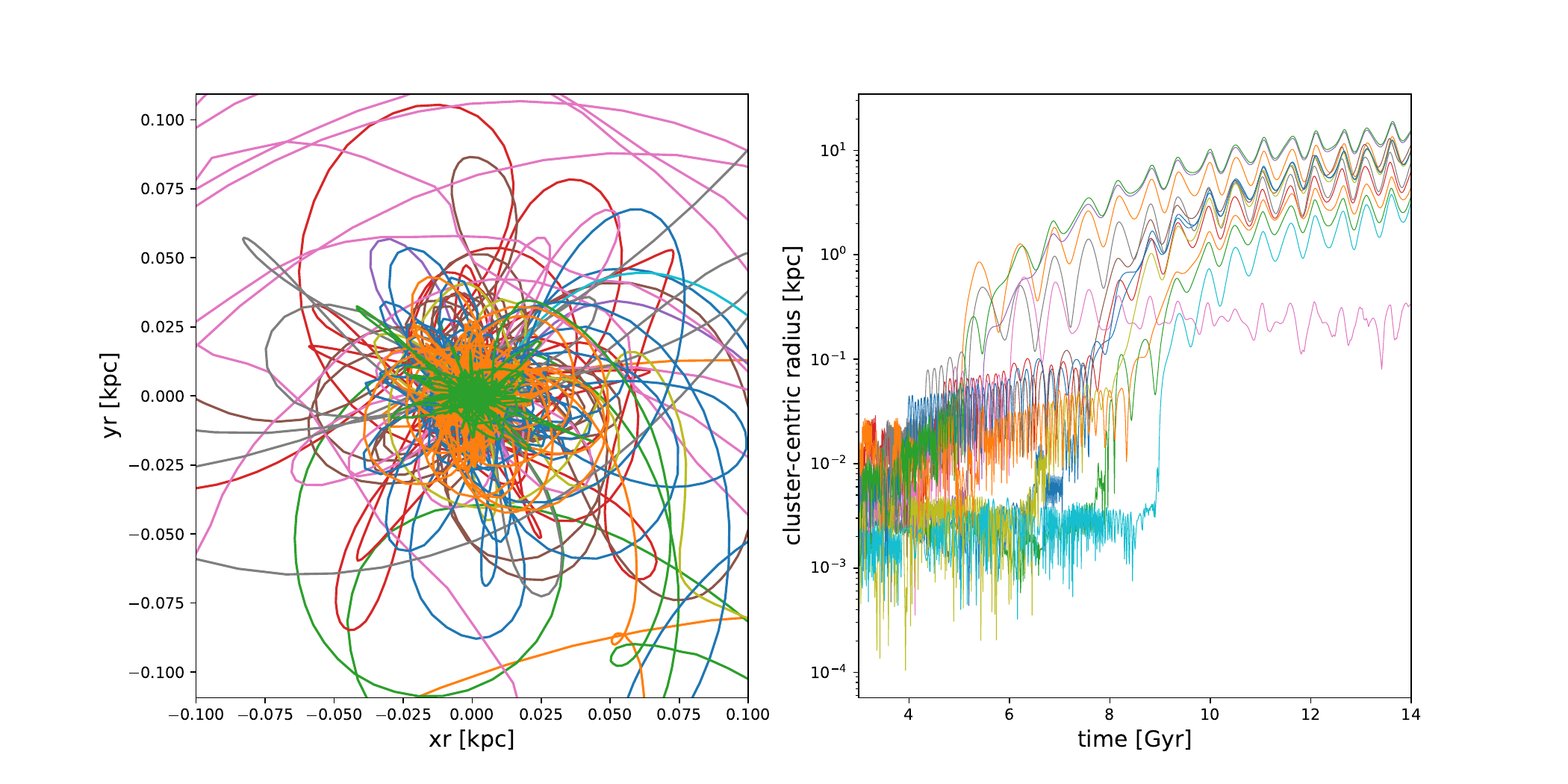}
\caption{Stars orbiting in the local rotating xy frame with x away from the galactic center (left) and the radius with time for the same stars (right). The stream is oriented vertically at larger distances.}
\label{fig_rotxyr}
\end{figure*}

\begin{figure}
\includegraphics[scale=0.55,trim=0 20 0 20, clip=true]{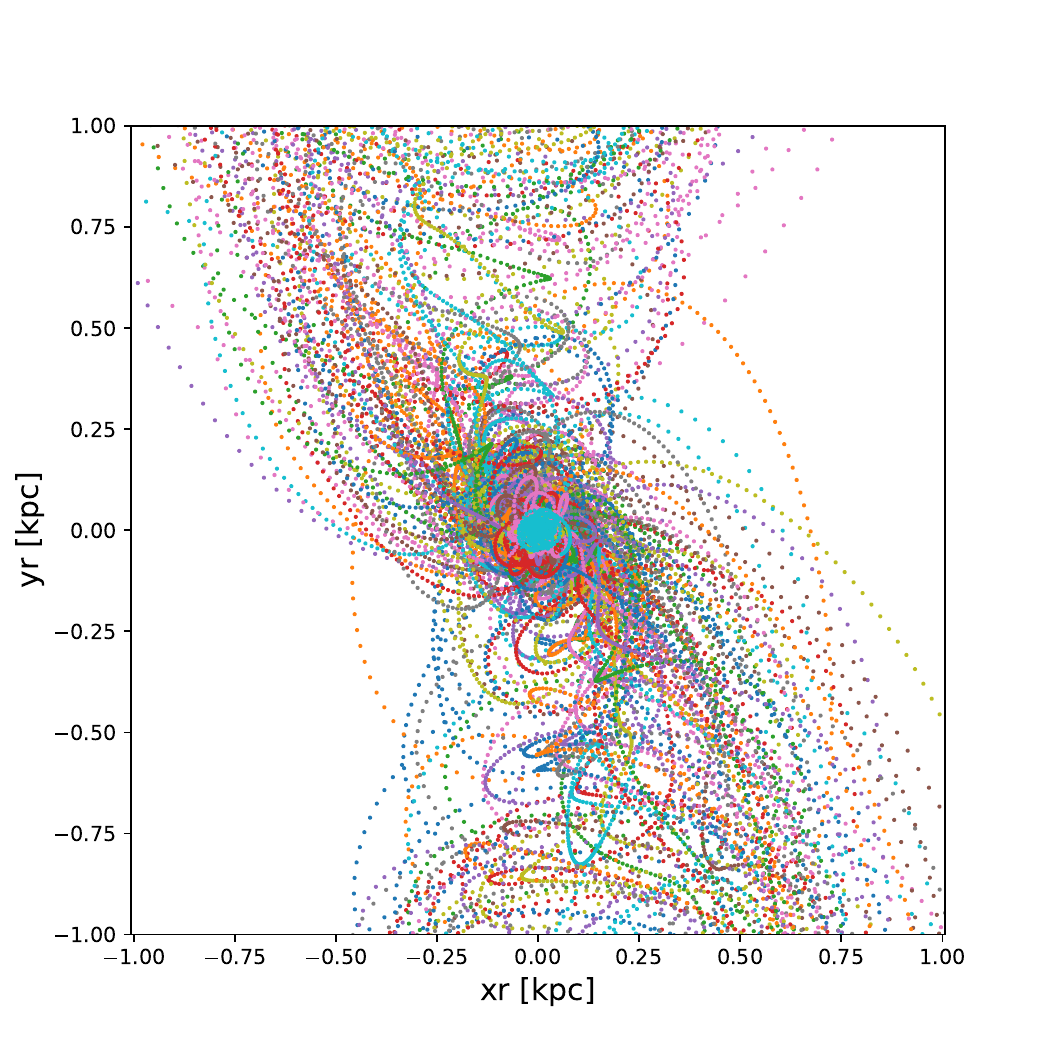}
\caption{Star particle orbits in the same rotating xy frame as Figure~\ref{fig_rotxyr} into the stream.}
\label{fig_tidalorb}
\end{figure}

{ The angular momentum and energy spread of stars in the stream is important for stream structure and affects the stream's response to subhalos. The evolution of a $5\times 10^4 M_\odot$ cluster started at 3 Gyr with the CDM subhalos at that time is shown in Figure~\ref{fig_cluster}. Cluster quantities are measured for particles within 20 pc of the center to discard particles leaving for the stream. The cluster is effectively unbound at 11.5 Gyr, but a group of particles near the mean halo density continue to orbit together. The star particle cluster velocity dispersion drops smoothly from an initial value around 3.5 \kms\ to close to zero at the time of dissolution. The orbits of a few  star particles inside the tidal region, approximately 100 pc in radius, are shown in the left panel of Figure~\ref{fig_rotxyr} and their radial evolution shown in the right panel. The orbits are complex \citep{FH00} and are tidally heated at pericenter passages gradually rising out to the tidal radius \citep{BT08,Meiron21}. The orbits are in rotating frame with the x axis pointed away from galactic center and y in the local orbital plane. The Monte Carlo heating of the particles operates within the inner 6 pc, primarly serving to boost particles into the tidal zone where the tidal field of the galaxy-cluster combination control their orbital evolution. Particles in the 1 kpc region are shown in Figure~\ref{fig_tidalorb}.. The stream has a spread in energy and angular momentum which blurs subhalo induced velocity changes and spreads them along the stream. }

\begin{figure}
\includegraphics[scale=0.75,trim=0 0 0 30, clip=true]{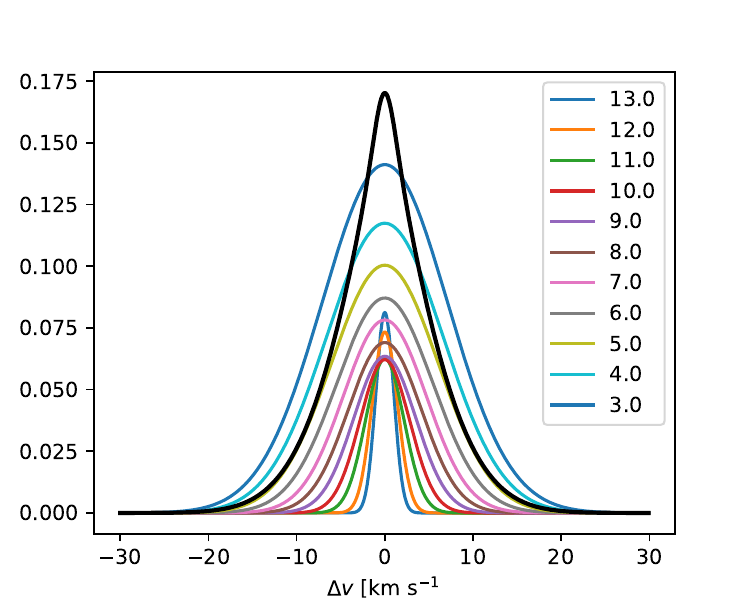}
\caption{A simple model for the velocity distribution for the CDM heating of Figure~\ref{fig_heating} with cluster mass loss into the stream declining as an exponential with a time constant of 4 Gyr. The colored lines are the distributions for particles entering the stream with time. The legend gives the time when the stars entered the stream, with the present being 14 Gyr.  The black line is the resulting overall velocity distribution, with a total velocity dispersion of 5.7 \kms\ and 3 and 2 sigma clipped velocity dispersions of 5.5 and 4.3 \kms.}
\label{fig_vdist}
\end{figure}

The subhalo heating model predicts a velocity distribution function. If the star cluster quickly dissolved in the first Gyr all the star particles would be heated equally to about 7.7 \kms\ in a Gaussian distribution, assumed on the basis of random walk heating. However the particles continue to be released over time, with the cluster reaching 1\% of its initial mass at 10 Gyr. The more recently released particles are heated less so the combined velocity distribution is a sum of Gaussians of decreasing width, as shown in Figure~\ref{fig_vdist}. For this figure the mass declines exponentially with a time constant of 4 Gyr. The resulting distribution has a standard deviation of 5.72 \kms, with a 3 and 2$\sigma$ clipped standard deviations of 5.49 and 4.29 \kms. The limit case of a constant mass loss rate gives 4.55 \kms, with 3 and 2$\sigma$ standard deviations of 4.10 and 2.98 \kms.

\begin{figure}
\includegraphics[scale=0.33,trim=50 40 0 25, clip=true]{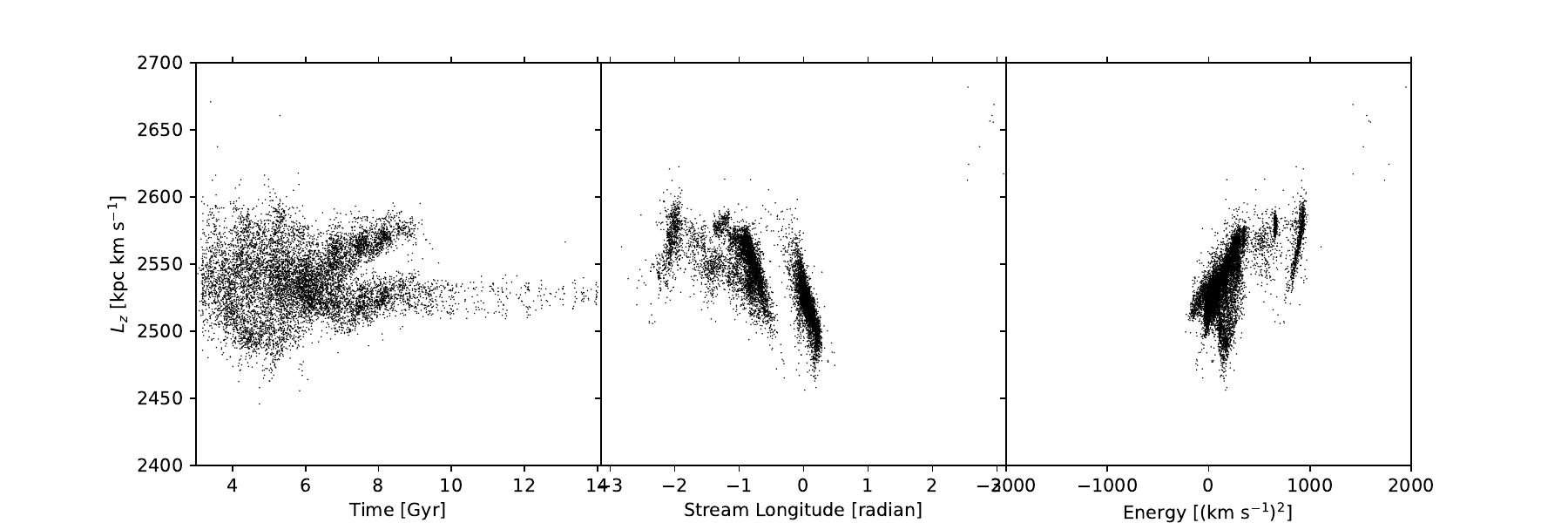}
\includegraphics[scale=0.33,trim=50 40 0 25, clip=true]{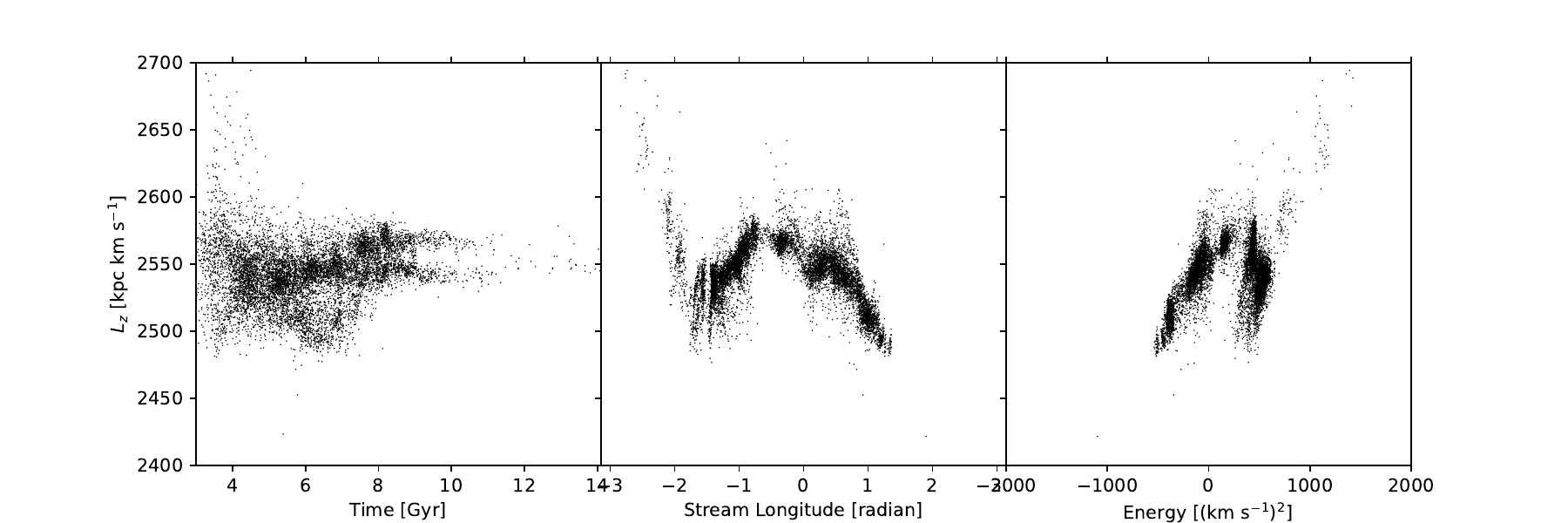}
\includegraphics[scale=0.33,trim=50 40  0 25, clip=true]{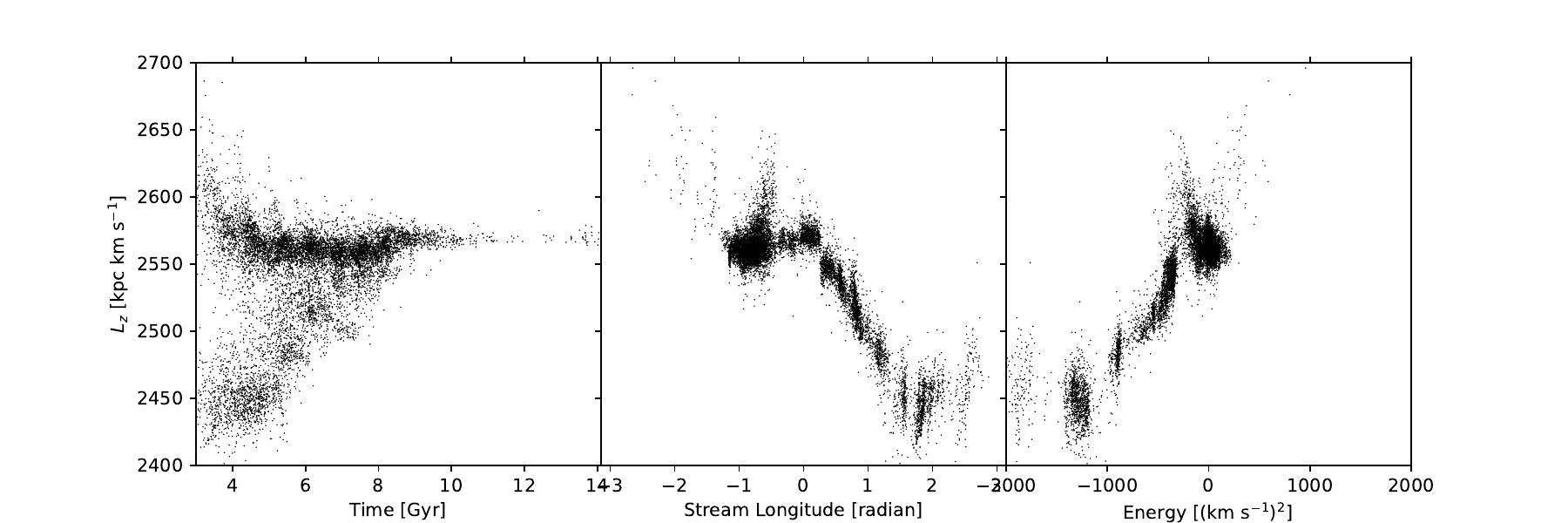}
\includegraphics[scale=0.33,trim=50   0 0 25, clip=true]{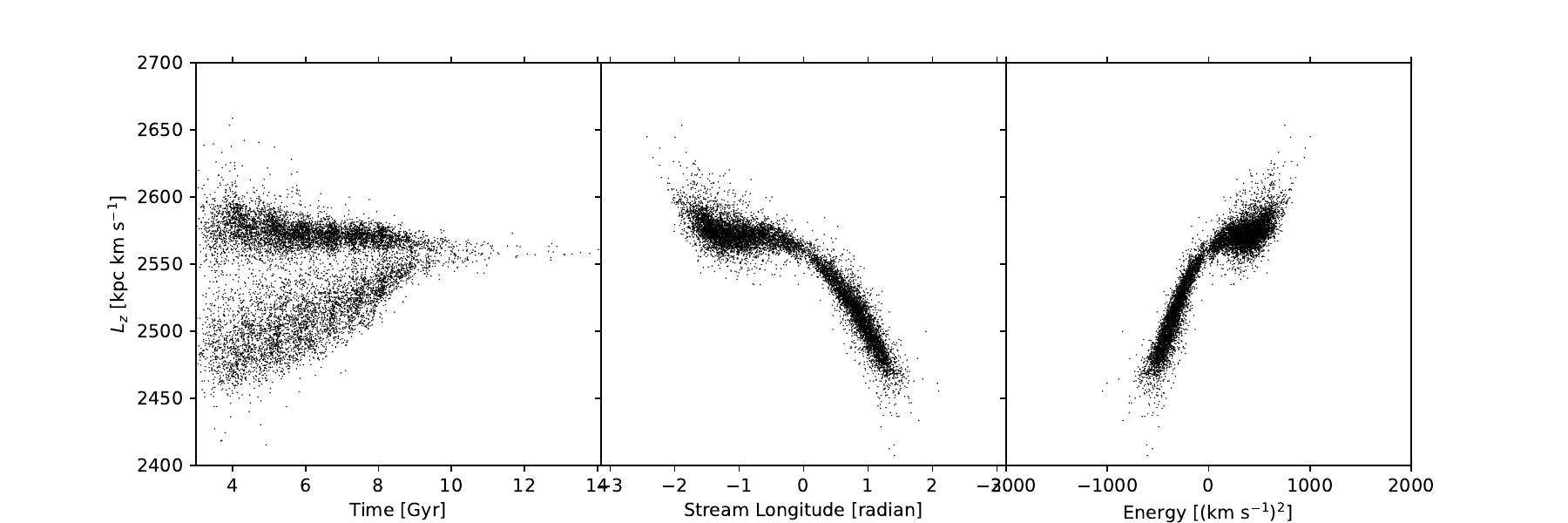}
\caption{The final time $L_z$ as a function of time when the particles are released (left), the galactocentric great circle longitude (center) and energy (right), where the progenitor center defines the zero of longitude and energy. Every fifth particle is plotted. The three top panels are simulations with the CDM subhalo populations rotated 0, 30 and 60 degrees, but are otherwise identical. The bottom panel is for a simulation with no subhalos and no dwarf galaxies.}
\label{fig_Lz}
\end{figure}

The heating process is somewhat nonlinear as illustrated in Figure~\ref{fig_Lz}. The panels on the left show the final time $L_z$ as a function of the time at which the particles (one in five) were at least 20 pc away from the cluster center. The middle panels show $L_z$ as a function of the galactocentric great circle longitude relative to the progenitor center. The right panels display $L_z$ against the particle energy calculated from the model potential at the final time, measured relative to the energy of the central star particle. The bottom two panels are for a simulation with no subhalos and no dwarf galaxies.  The upper three set of panels show three CDM simulations, the same as those in the first three (g282-4) of the top panels of Figure~\ref{fig_cdm11}, which in the top left are from the bottom, the blue, green and red points. The three CDM simulations use exactly the same galactic potential, with the subhalo population rotated 0, 30 and 60 degrees. These figures underscore the highly stochastic nature of the heating and that the most massive subhalos have few encounters but have large effects leading to the larger gaps. The longitude distribution of the middle panel shows that particles are moved in stream longitude to create gaps. The relative movement of orbital pericenters leads to an increase of velocity dispersion along the stream, with little or no change in the orbital energy or $L_z$ of the stream particles.   Movies showing these simulations are available at  \href{https://www.astro.utoronto.ca/~carlberg/streams/gd1}{GD-1 simulations}. 

In this paper the models are compared to the velocities aggregated in the $\phi_1=$[-30, 0] region where the stream is best sampled. The width of GD-1 and the density fluctuations along its length increase in proportion to the numbers of subhalos and are also important measures of the subhalo population. These quantities and the density variations will be considered in future papers as the sample size grows and the sampling function along the stream is better quantified.

\begin{figure}
\includegraphics[scale=0.23,trim=0 0 10 0, clip=true]{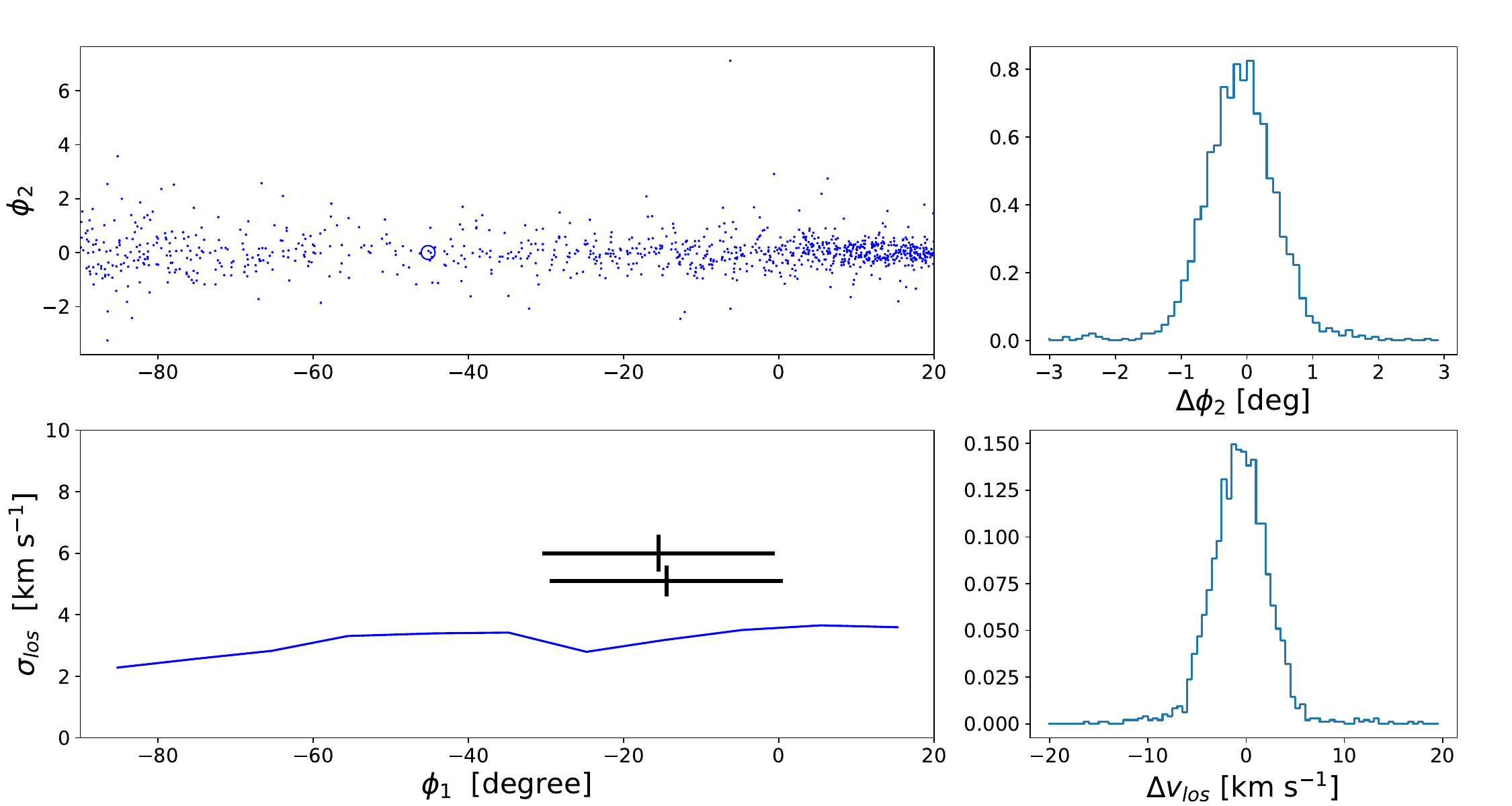}
\includegraphics[scale=0.23,trim=0 0 10 0, clip=true]{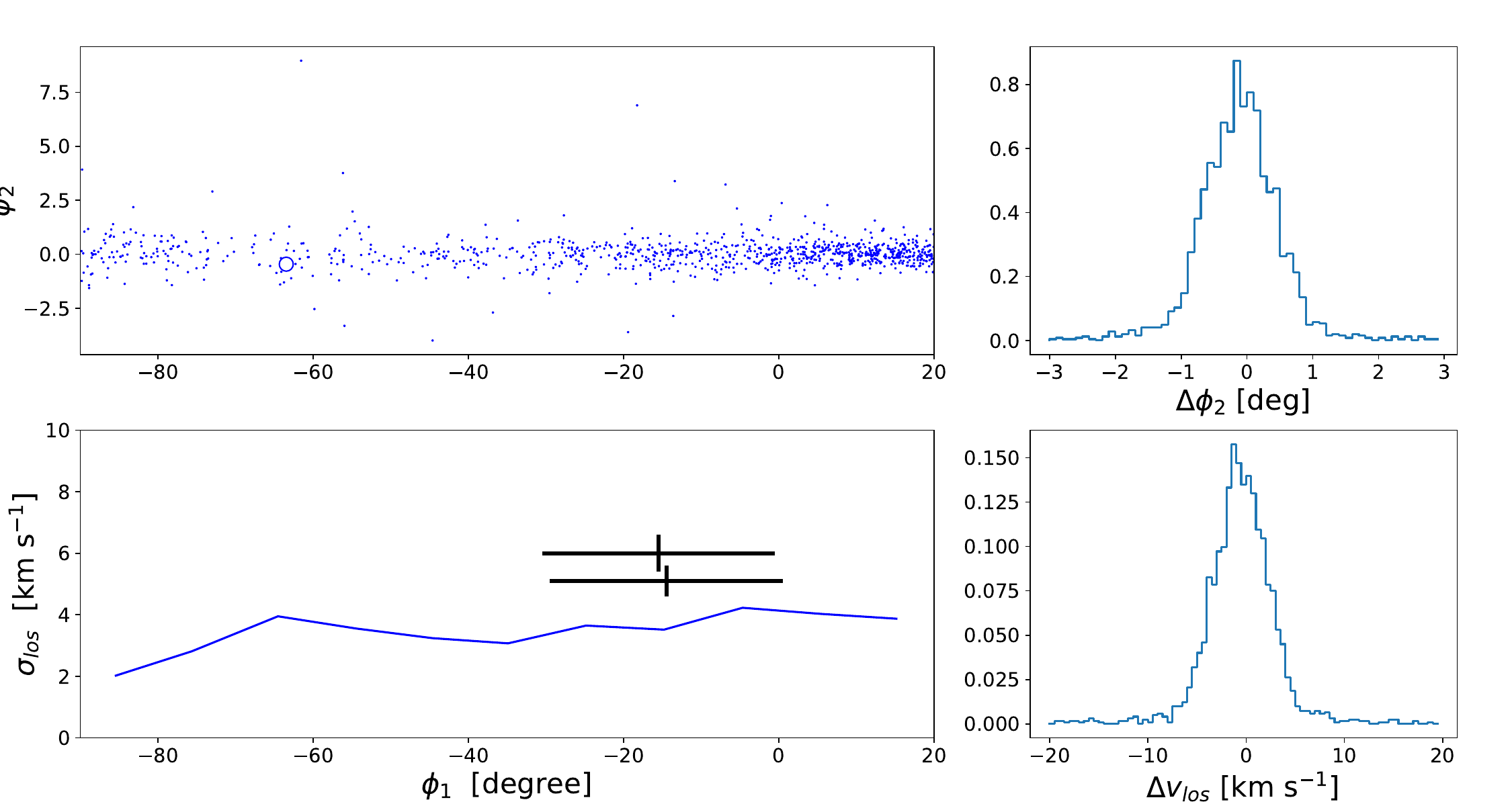}
\caption{A stream from a $5\times10^4 M_\odot$ progenitor in a halo with no subhalos and no dwarf galaxies (top) and dwarfs only (bottom). 
} 
\label{fig_nono}
\end{figure}

\section{Discussion and Conclusions\label{sec_conc}}

Measurement of the velocity dispersion of the stars to nearly 3 degrees from the stream centerline reveals that  GD-1 stream velocity distribution has a non-Gaussian velocity distribution. The standard deviation of the velocities in this region is  6.0$\pm$ 0.6 \kms\ and the 3$\sigma$ clipped velocity dispersion is 5.1 $\pm$ 0.5 \kms\ (the same as the $\left| \Delta v \right| < 20$ \kms\ sample) after removing the RMS velocity errors in quadrature. These velocity widths cannot be produced for a tidal stream from a globular cluster orbiting in a smooth Milky Way potential.  

The n-body simulations of GD-1 presented here use an evolving primary halo potential and an evolving subhalo population. The primary Milky Way halo grows in mass with time in accord with a cosmological simulation approximately matched to the Milky Way, but the model potential has a fixed triaxial shape, as is likely appropriate for an inner halo orbit. A population of subhalos is drawn from the same cosmological simulation and their orbits calculated in the primary potential. The subhalo masses are reduced with time to account for their evolution. The model potential allows individual streams, such as GD-1, to be evolved to their current locations in an evolving, cosmologically motivated potential. That is, it is designed to approximate the overall galactic scale, 30 kpc,  potential  and the small scale, 1 kpc, subhalos, but it does not account for the fraying of the ends of the stream caused by galactic scale merger activity that occurs in fully cosmological stream simulations. 

Dynamical modeling of progenitor globular clusters of  mass $2\times 10^4 M_\odot$ and 5$\times 10^4 M_\odot$ in a Milky Way potential containing subhalos drawn from cosmological simulations of Milky Way-like galaxies leads to results with $\simeq 3\sigma$ significance and some stream membership uncertainties.
The known dwarf galaxies do not provide any substantial heating to the GD-1 stream \citep{Bonaca19}, even for a stream age of 11 Gyr.
There are insufficient dark subhalos in 7 keV warm dark matter (and any lower mass value)  to heat the stream for stream ages of 11 Gyr or less for the same orbital parameters as the CDM models.  A wider range of  models should be examined.  
The CDM subhalo population for streams 9 Gyr and younger is insufficient to heat the stream.
The evolving CDM subhalo population, which has about a factor of 5 more subhalos at early times,  heats the GD-1 stream to the observed width and velocity spread  for a cluster started at 3 Gyr  and allowed to orbit to the current epoch, a total of about 11 Gyr.
 The best model for the velocity width of GD-1 requires a  progenitor star cluster of mass  $5\times 10^4 M_\odot$. About 60\% of its stars are dispersed over a wider stream around the galaxy, leaving about $2\times 10^4 M_\odot$ in the visible high galactic latitude stream. The evolving potential model does not accurately predict the dispersal of the ends of the stream that are seen in the cosmological stream n-body simulations. 

The models have density variations along their streams that correlate with the rise in the velocity spread in the stream. The current data has significant sampling variations and is not well suited for comparison to the stream density variation models. The model density variations appear similar to those of Figure~3 of \citep{Tavangar25}.

{ The GD-1 stream has received considerable attention for its ability to constrain the Milky Way subhalo population \citep{Bonaca19,B21}. An important outcome is the consensus view that baryonic structures of the Milky Way, such as molecular clouds, the galactic bar, and the disk itself are not sufficient to account for the structure of GD-1 \citep{B21}.   \citet{M21,M22} have shown that an accreted dwarf galaxy containing a globular cluster can produce streams with a narrow component from the cluster and a wider component from the stars lost while the cluster orbits within the dwarf galaxy.  The star clusters in the cosmological n-body simulations here were inserted into early time halos with that had dwarf galaxy scale masses, although they also have internal subhalos \citep{Springel08}. As in Figure 4 of \citet{M21}, our Figure~\ref{fig_siglen} shows that the pre-accretion cluster stars near the progenitor cluster stay together in a stream that varies in velocity dispersion and width as it orbits. However the stream in the dwarf has stars spread around an orbit which when disrupted are widely scattered over the halo in the complex potential of the complex, time varying, potential of cosmological simulation \citep{CA,Carlberg24}. The movies at \href{https://www.astro.utoronto.ca/~carlberg/streams/417}{CDM Low Mass Cluster Streams} show both the entire simulation and several individual streams.  } 

The results are entirely dependent on the subhalos within the Milky Way like potential, as motivated by cosmological simulations. The primary halo selected for detailed simulation builds up rapidly and has its last major merger about 7 Gyr in the past, with ongoing accretion. A Miyamoto-Nagai bulge-disk with MW2014 parameters \citep{galpy} builds its mass from zero at 5 Gyr to full mass at the end of the  simulation. Together the declining number of infalling new subhalos and increasing central tidal fields cause the  subhalo population to drop substantially with time. Modeling of C-19 in this potential found similar results \citep{C19Carlberg} to those presented.  Two versions of the progenitor star cluster are explored here. A wider range of masses, initial half mass radii, and orbital positioning will change stream details which will be important as the wide area sampling of GD-1 increases. 

\begin{acknowledgements}
I thank the referee for an insightful review that helped improve this paper. 
Computations were performed on the niagara supercomputer at the SciNet HPC Consortium. SciNet is funded by: the Canada Foundation for Innovation; the Government of Ontario; Ontario Research Fund - Research Excellence; and the University of Toronto. 
\end{acknowledgements}

\software{Gadget4: \citet{Gadget4}, Amiga Halo Finder: \citep{AHF1,AHF2}}, NumPy: \citep{numpy}.

Data Availability: Final snapshots, movies, images, and example scripts are at \href{https://www.astro.utoronto.ca/~carlberg/streams}{Streams Data} and its subdirectories.

\bibliography{GD1}{}

\begin{thebibliography}{}
\expandafter\ifx\csname natexlab\endcsname\relax\def\natexlab#1{#1}\fi
\providecommand{\url}[1]{\href{#1}{#1}}
\providecommand{\dodoi}[1]{doi:~\href{http://doi.org/#1}{\nolinkurl{#1}}}
\providecommand{\doeprint}[1]{\href{http://ascl.net/#1}{\nolinkurl{http://ascl.net/#1}}}
\providecommand{\doarXiv}[1]{\href{https://arxiv.org/abs/#1}{\nolinkurl{https://arxiv.org/abs/#1}}}

\bibitem[{{Banik} {et~al.}(2021){Banik}, {Bovy}, {Bertone}, {Erkal}, \& {de
  Boer}}]{B21}
{Banik}, N., {Bovy}, J., {Bertone}, G., {Erkal}, D., \& {de Boer}, T.~J.~L.
  2021, \mnras, 502, 2364, \dodoi{10.1093/mnras/stab210}

\bibitem[{{Belokurov} {et~al.}(2006){Belokurov}, {Zucker}, {Evans}, {Gilmore},
  {Vidrih}, {Bramich}, {Newberg}, {Wyse}, {Irwin}, {Fellhauer}, {Hewett},
  {Walton}, {Wilkinson}, {Cole}, {Yanny}, {Rockosi}, {Beers}, {Bell},
  {Brinkmann}, {Ivezi{\'c}}, \& {Lupton}}]{g06}
{Belokurov}, V., {Zucker}, D.~B., {Evans}, N.~W., {et~al.} 2006, \apjl, 642,
  L137, \dodoi{10.1086/504797}

\bibitem[{{Benitez-Llambay} \& {Frenk}(2020)}]{Benitez-Llambay20}
{Benitez-Llambay}, A., \& {Frenk}, C. 2020, \mnras, 498, 4887,
  \dodoi{10.1093/mnras/staa2698}

\bibitem[{{Binney} \& {Tremaine}(2008)}]{BT08}
{Binney}, J., \& {Tremaine}, S. 2008, {Galactic Dynamics: Second Edition}
  (Princeton University Press)

\bibitem[{{Bode} {et~al.}(2001){Bode}, {Ostriker}, \& {Turok}}]{Bode01}
{Bode}, P., {Ostriker}, J.~P., \& {Turok}, N. 2001, \apj, 556, 93,
  \dodoi{10.1086/321541}

\bibitem[{{Bonaca} {et~al.}(2019){Bonaca}, {Hogg}, {Price-Whelan}, \&
  {Conroy}}]{Bonaca19}
{Bonaca}, A., {Hogg}, D.~W., {Price-Whelan}, A.~M., \& {Conroy}, C. 2019, \apj,
  880, 38, \dodoi{10.3847/1538-4357/ab2873}

\bibitem[{{Bonaca} \& {Price-Whelan}(2024)}]{BPW24}
{Bonaca}, A., \& {Price-Whelan}, A.~M. 2024, arXiv e-prints, arXiv:2405.19410,
  \dodoi{10.48550/arXiv.2405.19410}

\bibitem[{{Bovy}(2015)}]{galpy}
{Bovy}, J. 2015, \apjs, 216, 29, \dodoi{10.1088/0067-0049/216/2/29}

\bibitem[{{Carlberg}(2009)}]{Carlberg09}
{Carlberg}, R.~G. 2009, \apjl, 705, L223, \dodoi{10.1088/0004-637X/705/2/L223}

\bibitem[{{Carlberg}(2013)}]{Carlberg13}
---. 2013, \apj, 775, 90, \dodoi{10.1088/0004-637X/775/2/90}

\bibitem[{{Carlberg}(2018)}]{Carlberg18}
---. 2018, \apj, 861, 69, \dodoi{10.3847/1538-4357/aac88a}

\bibitem[{{Carlberg} \& {Agler}(2023)}]{CA23}
{Carlberg}, R.~G., \& {Agler}, H. 2023, \apj, 953, 99,
  \dodoi{10.3847/1538-4357/ace4be}

\bibitem[{{Carlberg} {et~al.}(2024{\natexlab{a}}){Carlberg}, {Ibata}, {Martin},
  {Starkenburg}, {Aguado}, {Malhan}, {Venn}, \& {Venn}}]{C19Carlberg}
{Carlberg}, R.~G., {Ibata}, R., {Martin}, N.~F., {et~al.} 2024{\natexlab{a}},
  arXiv e-prints, arXiv:2410.22966, \dodoi{10.48550/arXiv.2410.22966}

\bibitem[{{Carlberg} {et~al.}(2024{\natexlab{b}}){Carlberg}, {Jenkins},
  {Frenk}, \& {Cooper}}]{Carlberg24}
{Carlberg}, R.~G., {Jenkins}, A., {Frenk}, C.~S., \& {Cooper}, A.~P.
  2024{\natexlab{b}}, arXiv e-prints, arXiv:2405.18522,
  \dodoi{10.48550/arXiv.2405.18522}

\bibitem[{{Cooper} {et~al.}(2023){Cooper}, {Koposov}, {Allende Prieto},
  {Manser}, {Kizhuprakkat}, {Myers}, {Dey}, {G{\"a}nsicke}, {Li}, {Rockosi},
  {Valluri}, {Najita}, {Deason}, {Raichoor}, {Wang}, {Ting}, {Kim}, {Carrillo},
  {Wang}, {Beraldo e Silva}, {Han}, {Ding}, {S{\'a}nchez-Conde}, {Aguilar},
  {Ahlen}, {Bailey}, {Belokurov}, {Brooks}, {Cunha}, {Dawson}, {de la Macorra},
  {Doel}, {Eisenstein}, {Fagrelius}, {Fanning}, {Font-Ribera}, {Forero-Romero},
  {Gazta{\~n}aga}, {Gontcho a Gontcho}, {Guy}, {Honscheid}, {Kehoe}, {Kisner},
  {Kremin}, {Landriau}, {Levi}, {Martini}, {Meisner}, {Miquel}, {Moustakas},
  {Nie}, {Palanque-Delabrouille}, {Percival}, {Poppett}, {Prada}, {Rehemtulla},
  {Schlafly}, {Schlegel}, {Schubnell}, {Sharples}, {Tarl{\'e}}, {Wechsler},
  {Weinberg}, {Zhou}, \& {Zou}}]{DESI-MWS}
{Cooper}, A.~P., {Koposov}, S.~E., {Allende Prieto}, C., {et~al.} 2023, \apj,
  947, 37, \dodoi{10.3847/1538-4357/acb3c0}

\bibitem[{{DESI Collaboration} {et~al.}(2016){DESI Collaboration}, {Aghamousa},
  {Aguilar}, {Ahlen}, {Alam}, {Allen}, {Allende Prieto}, {Annis}, {Bailey},
  {Balland}, {Ballester}, {Baltay}, {Beaufore}, {Bebek}, {Beers}, {Bell},
  {Bernal}, {Besuner}, {Beutler}, {Blake}, {Bleuler}, {Blomqvist}, {Blum},
  {Bolton}, {Briceno}, {Brooks}, {Brownstein}, {Buckley-Geer}, {Burden},
  {Burtin}, {Busca}, {Cahn}, {Cai}, {Cardiel-Sas}, {Carlberg}, {Carton},
  {Casas}, {Castander}, {Cervantes-Cota}, {Claybaugh}, {Close}, {Coker},
  {Cole}, {Comparat}, {Cooper}, {Cousinou}, {Crocce}, {Cuby}, {Cunningham},
  {Davis}, {Dawson}, {de la Macorra}, {De Vicente}, {Delubac}, {Derwent},
  {Dey}, {Dhungana}, {Ding}, {Doel}, {Duan}, {Ealet}, {Edelstein},
  {Eftekharzadeh}, {Eisenstein}, {Elliott}, {Escoffier}, {Evatt}, {Fagrelius},
  {Fan}, {Fanning}, {Farahi}, {Farihi}, {Favole}, {Feng}, {Fernandez},
  {Findlay}, {Finkbeiner}, {Fitzpatrick}, {Flaugher}, {Flender}, {Font-Ribera},
  {Forero-Romero}, {Fosalba}, {Frenk}, {Fumagalli}, {Gaensicke}, {Gallo},
  {Garcia-Bellido}, {Gaztanaga}, {Pietro Gentile Fusillo}, {Gerard},
  {Gershkovich}, {Giannantonio}, {Gillet}, {Gonzalez-de-Rivera},
  {Gonzalez-Perez}, {Gott}, {Graur}, {Gutierrez}, {Guy}, {Habib}, {Heetderks},
  {Heetderks}, {Heitmann}, {Hellwing}, {Herrera}, {Ho}, {Holland}, {Honscheid},
  {Huff}, {Hutchinson}, {Huterer}, {Hwang}, {Illa Laguna}, {Ishikawa},
  {Jacobs}, {Jeffrey}, {Jelinsky}, {Jennings}, {Jiang}, {Jimenez}, {Johnson},
  {Joyce}, {Jullo}, {Juneau}, {Kama}, {Karcher}, {Karkar}, {Kehoe}, {Kennamer},
  {Kent}, {Kilbinger}, {Kim}, {Kirkby}, {Kisner}, {Kitanidis}, {Kneib},
  {Koposov}, {Kovacs}, {Koyama}, {Kremin}, {Kron}, {Kronig}, {Kueter-Young},
  {Lacey}, {Lafever}, {Lahav}, {Lambert}, {Lampton}, {Landriau}, {Lang},
  {Lauer}, {Le Goff}, {Le Guillou}, {Le Van Suu}, {Lee}, {Lee}, {Leitner},
  {Lesser}, {Levi}, {L'Huillier}, {Li}, {Liang}, {Lin}, {Linder}, {Loebman},
  {Luki{\'c}}, {Ma}, {MacCrann}, {Magneville}, {Makarem}, {Manera}, {Manser},
  {Marshall}, {Martini}, {Massey}, {Matheson}, {McCauley}, {McDonald},
  {McGreer}, {Meisner}, {Metcalfe}, {Miller}, {Miquel}, {Moustakas}, {Myers},
  {Naik}, {Newman}, {Nichol}, {Nicola}, {Nicolati da Costa}, {Nie}, {Niz},
  {Norberg}, {Nord}, {Norman}, {Nugent}, {O'Brien}, {Oh}, \&
  {Olsen}}]{DESICollab}
{DESI Collaboration}, {Aghamousa}, A., {Aguilar}, J., {et~al.} 2016, arXiv
  e-prints, arXiv:1611.00036, \dodoi{10.48550/arXiv.1611.00036}

\bibitem[{{D'Onghia} {et~al.}(2010){D'Onghia}, {Springel}, {Hernquist}, \&
  {Keres}}]{Donghia10}
{D'Onghia}, E., {Springel}, V., {Hernquist}, L., \& {Keres}, D. 2010, \apj,
  709, 1138, \dodoi{10.1088/0004-637X/709/2/1138}

\bibitem[{{Erkal} \& {Belokurov}(2015)}]{Erkal15}
{Erkal}, D., \& {Belokurov}, V. 2015, \mnras, 454, 3542,
  \dodoi{10.1093/mnras/stv2122}

\bibitem[{{Fukushige} \& {Heggie}(2000)}]{FH00}
{Fukushige}, T., \& {Heggie}, D.~C. 2000, \mnras, 318, 753,
  \dodoi{10.1046/j.1365-8711.2000.03811.x}

\bibitem[{{Garrison-Kimmel} {et~al.}(2017){Garrison-Kimmel}, {Wetzel},
  {Bullock}, {Hopkins}, {Boylan-Kolchin}, {Faucher-Gigu{\`e}re}, {Kere{\v{s}}},
  {Quataert}, {Sanderson}, {Graus}, \& {Kelley}}]{GKBullock17}
{Garrison-Kimmel}, S., {Wetzel}, A., {Bullock}, J.~S., {et~al.} 2017, \mnras,
  471, 1709, \dodoi{10.1093/mnras/stx1710}

\bibitem[{{Gill} {et~al.}(2004){Gill}, {Knebe}, \& {Gibson}}]{AHF1}
{Gill}, S. P.~D., {Knebe}, A., \& {Gibson}, B.~K. 2004, \mnras, 351, 399,
  \dodoi{10.1111/j.1365-2966.2004.07786.x}

\bibitem[{{Grillmair} \& {Dionatos}(2006)}]{GD1}
{Grillmair}, C.~J., \& {Dionatos}, O. 2006, \apjl, 643, L17,
  \dodoi{10.1086/505111}

\bibitem[{{Grillmair} {et~al.}(1995){Grillmair}, {Freeman}, {Irwin}, \&
  {Quinn}}]{Grillmair95}
{Grillmair}, C.~J., {Freeman}, K.~C., {Irwin}, M., \& {Quinn}, P.~J. 1995, \aj,
  109, 2553, \dodoi{10.1086/117470}

\bibitem[{Harris {et~al.}(2020)Harris, Millman, van~der Walt, Gommers,
  Virtanen, Cournapeau, Wieser, Taylor, Berg, Smith, Kern, Picus, Hoyer, van
  Kerkwijk, Brett, Haldane, del R{\'{i}}o, Wiebe, Peterson,
  G{\'{e}}rard-Marchant, Sheppard, Reddy, Weckesser, Abbasi, Gohlke, \&
  Oliphant}]{numpy}
Harris, C.~R., Millman, K.~J., van~der Walt, S.~J., {et~al.} 2020, Nature, 585,
  357, \dodoi{10.1038/s41586-020-2649-2}

\bibitem[{{Ibata} {et~al.}(2024){Ibata}, {Malhan}, {Tenachi},
  {Ardern-Arentsen}, {Bellazzini}, {Bianchini}, {Bonifacio}, {Caffau},
  {Diakogiannis}, {Errani}, {Famaey}, {Ferrone}, {Martin}, {di Matteo},
  {Monari}, {Renaud}, {Starkenburg}, {Thomas}, {Viswanathan}, \&
  {Yuan}}]{Ibata24}
{Ibata}, R., {Malhan}, K., {Tenachi}, W., {et~al.} 2024, \apj, 967, 89,
  \dodoi{10.3847/1538-4357/ad382d}

\bibitem[{{Ibata} {et~al.}(2002){Ibata}, {Lewis}, {Irwin}, \&
  {Quinn}}]{Ibata02}
{Ibata}, R.~A., {Lewis}, G.~F., {Irwin}, M.~J., \& {Quinn}, T. 2002, \mnras,
  332, 915, \dodoi{10.1046/j.1365-8711.2002.05358.x}

\bibitem[{{Johnston} {et~al.}(2002){Johnston}, {Spergel}, \&
  {Haydn}}]{Johnston02}
{Johnston}, K.~V., {Spergel}, D.~N., \& {Haydn}, C. 2002, \apj, 570, 656,
  \dodoi{10.1086/339791}

\bibitem[{{Klypin} {et~al.}(1999){Klypin}, {Kravtsov}, {Valenzuela}, \&
  {Prada}}]{Klypin99}
{Klypin}, A., {Kravtsov}, A.~V., {Valenzuela}, O., \& {Prada}, F. 1999, \apj,
  522, 82, \dodoi{10.1086/307643}

\bibitem[{{Knollmann} \& {Knebe}(2009)}]{AHF2}
{Knollmann}, S.~R., \& {Knebe}, A. 2009, \apjs, 182, 608,
  \dodoi{10.1088/0067-0049/182/2/608}

\bibitem[{{Koposov} {et~al.}(2010){Koposov}, {Rix}, \& {Hogg}}]{Koposov10}
{Koposov}, S.~E., {Rix}, H.-W., \& {Hogg}, D.~W. 2010, \apj, 712, 260,
  \dodoi{10.1088/0004-637X/712/1/260}

\bibitem[{{Lovell} {et~al.}(2014){Lovell}, {Frenk}, {Eke}, {Jenkins}, {Gao}, \&
  {Theuns}}]{Lovell14}
{Lovell}, M.~R., {Frenk}, C.~S., {Eke}, V.~R., {et~al.} 2014, \mnras, 439, 300,
  \dodoi{10.1093/mnras/stt2431}

\bibitem[{{Malhan} \& {Ibata}(2018)}]{Malhan18}
{Malhan}, K., \& {Ibata}, R.~A. 2018, \mnras, 477, 4063,
  \dodoi{10.1093/mnras/sty912}

\bibitem[{{Malhan} {et~al.}(2019){Malhan}, {Ibata}, {Carlberg}, {Valluri}, \&
  {Freese}}]{Malhan19}
{Malhan}, K., {Ibata}, R.~A., {Carlberg}, R.~G., {Valluri}, M., \& {Freese}, K.
  2019, \apj, 881, 106, \dodoi{10.3847/1538-4357/ab2e07}

\bibitem[{{Malhan} {et~al.}(2021){Malhan}, {Valluri}, \& {Freese}}]{M21}
{Malhan}, K., {Valluri}, M., \& {Freese}, K. 2021, \mnras, 501, 179,
  \dodoi{10.1093/mnras/staa3597}

\bibitem[{{Malhan} {et~al.}(2022){Malhan}, {Valluri}, {Freese}, \&
  {Ibata}}]{M22}
{Malhan}, K., {Valluri}, M., {Freese}, K., \& {Ibata}, R.~A. 2022, \apjl, 941,
  L38, \dodoi{10.3847/2041-8213/aca6e5}

\bibitem[{{Mateu}(2023)}]{Mateu23}
{Mateu}, C. 2023, \mnras, 520, 5225, \dodoi{10.1093/mnras/stad321}

\bibitem[{{McConnachie}(2012)}]{McConnachie12}
{McConnachie}, A.~W. 2012, \aj, 144, 4, \dodoi{10.1088/0004-6256/144/1/4}

\bibitem[{{Meiron} {et~al.}(2021){Meiron}, {Webb}, {Hong}, {Berczik},
  {Spurzem}, \& {Carlberg}}]{Meiron21}
{Meiron}, Y., {Webb}, J.~J., {Hong}, J., {et~al.} 2021, \mnras, 503, 3000,
  \dodoi{10.1093/mnras/stab649}

\bibitem[{{Miyamoto} \& {Nagai}(1975)}]{MN75}
{Miyamoto}, M., \& {Nagai}, R. 1975, \pasj, 27, 533

\bibitem[{{Moore} {et~al.}(1999){Moore}, {Ghigna}, {Governato}, {Lake},
  {Quinn}, {Stadel}, \& {Tozzi}}]{Moore99}
{Moore}, B., {Ghigna}, S., {Governato}, F., {et~al.} 1999, \apjl, 524, L19,
  \dodoi{10.1086/312287}

\bibitem[{{Navarro} {et~al.}(1996){Navarro}, {Frenk}, \& {White}}]{NFW}
{Navarro}, J.~F., {Frenk}, C.~S., \& {White}, S. D.~M. 1996, \apj, 462, 563,
  \dodoi{10.1086/177173}

\bibitem[{{Nibauer} {et~al.}(2024){Nibauer}, {Bonaca}, {Spergel},
  {Price-Whelan}, {Greene}, {Starkman}, \& {Johnston}}]{Nibauer24}
{Nibauer}, J., {Bonaca}, A., {Spergel}, D.~N., {et~al.} 2024, arXiv e-prints,
  arXiv:2410.21174, \dodoi{10.48550/arXiv.2410.21174}

\bibitem[{{Odenkirchen} {et~al.}(2001){Odenkirchen}, {Grebel}, {Rockosi},
  {Dehnen}, {Ibata}, {Rix}, {Stolte}, {Wolf}, {Anderson}, {Bahcall},
  {Brinkmann}, {Csabai}, {Hennessy}, {Hindsley}, {Ivezi{\'c}}, {Lupton},
  {Munn}, {Pier}, {Stoughton}, \& {York}}]{Odenkirchen01}
{Odenkirchen}, M., {Grebel}, E.~K., {Rockosi}, C.~M., {et~al.} 2001, \apjl,
  548, L165, \dodoi{10.1086/319095}

\bibitem[{{O'Riordan} \& {Vegetti}(2024)}]{Vegetti24}
{O'Riordan}, C.~M., \& {Vegetti}, S. 2024, \mnras, 528, 1757,
  \dodoi{10.1093/mnras/stae153}

\bibitem[{{Price-Whelan}(2017)}]{PW17}
{Price-Whelan}, A.~M. 2017, The Journal of Open Source Software, 2, 388,
  \dodoi{10.21105/joss.00388}

\bibitem[{{Sanders} \& {Binney}(2013)}]{SandersBinney13}
{Sanders}, J.~L., \& {Binney}, J. 2013, \mnras, 433, 1813,
  \dodoi{10.1093/mnras/stt806}

\bibitem[{Shen {et~al.}(2022)Shen, Eadie, Murray, Zaritsky, Speagle, Ting,
  Conroy, Cargile, Johnson, Naidu, \& Han}]{Shen22}
Shen, J., Eadie, G.~M., Murray, N., {et~al.} 2022, The Astrophysical Journal,
  925, 1, \dodoi{10.3847/1538-4357/ac3a7a}

\bibitem[{{Springel} {et~al.}(2021){Springel}, {Pakmor}, {Zier}, \&
  {Reinecke}}]{Gadget4}
{Springel}, V., {Pakmor}, R., {Zier}, O., \& {Reinecke}, M. 2021, \mnras, 506,
  2871, \dodoi{10.1093/mnras/stab1855}

\bibitem[{{Springel} {et~al.}(2008){Springel}, {Wang}, {Vogelsberger},
  {Ludlow}, {Jenkins}, {Helmi}, {Navarro}, {Frenk}, \& {White}}]{Springel08}
{Springel}, V., {Wang}, J., {Vogelsberger}, M., {et~al.} 2008, \mnras, 391,
  1685, \dodoi{10.1111/j.1365-2966.2008.14066.x}

\bibitem[{{Tavangar} \& {Price-Whelan}(2025)}]{Tavangar25}
{Tavangar}, K., \& {Price-Whelan}, A.~M. 2025, arXiv e-prints,
  arXiv:2502.13236, \dodoi{10.48550/arXiv.2502.13236}

\bibitem[{{Valluri} {et~al.}(2025){Valluri}, {Fagrelius}, {Koposov}, {Li},
  {Gnedin}, {Bell}, {Carlberg}, {Cooper}, {Aguilar}, {Ahlen}, {Allende Prieto},
  {Belokurov}, {Beraldo e Silva}, {Brooks}, {Bystr{\"o}m}, {Claybaugh},
  {Dawson}, {Dey}, {Doel}, {Forero-Romero}, {Gazta{\~n}aga}, {Gontcho A
  Gontcho}, {Han}, {Honscheid}, {Kisner}, {Kremin}, {Lambert}, {Landriau}, {Le
  Guillou}, {Levi}, {de la Macorra}, {Manera}, {Martini}, {Medina}, {Meisner},
  {Miquel}, {Moustakas}, {Myers}, {Najita}, {Poppett}, {Prada}, {Rezaie},
  {Rossi}, {Riley}, {Sanchez}, {Schlegel}, {Schubnell}, {Sprayberry},
  {Tarl{\'e}}, {Thomas}, {Weaver}, {Wechsler}, {Zhou}, \& {Zou}}]{Valluri25}
{Valluri}, M., {Fagrelius}, P., {Koposov}, S.~E., {et~al.} 2025, \apj, 980, 71,
  \dodoi{10.3847/1538-4357/ada690}

\bibitem[{{Webb} \& {Bovy}(2019)}]{WebbBovy19}
{Webb}, J.~J., \& {Bovy}, J. 2019, \mnras, 485, 5929,
  \dodoi{10.1093/mnras/stz867}

\end{thebibliography}
\bibliographystyle{aasjournal}

\end{document}